\documentclass[12pt]{article} 
\usepackage[colorlinks, citecolor={blue}]{hyperref}
\usepackage{url}
\usepackage{amsfonts,amscd,amssymb}
\usepackage{amsthm,amsmath,natbib}
\usepackage{algorithm,algorithmicx,algpseudocode}
\usepackage{bm}
\usepackage{bbm} 
\usepackage[table]{xcolor}
\usepackage{verbatim}
\usepackage{graphicx}
\usepackage{setspace}
\usepackage{natbib}
\usepackage[margin=1in]{geometry}
\usepackage{enumitem}
\usepackage{listings}
\usepackage[frozencache,cachedir=.]{minted}
\usepackage[textsize=tiny]{todonotes}
\usepackage{tikz}
\usepackage{etoolbox}
\usepackage{appendix}
\usepackage{subcaption}
\usepackage{wrapfig}
\usepackage{xr}
\usepackage{booktabs}
\usepackage{multirow}
\usepackage{authblk}

\usetikzlibrary{matrix}
\usetikzlibrary{backgrounds}
\usetikzlibrary{calc}
\usetikzlibrary{arrows,shapes}
\usetikzlibrary{decorations}
\usetikzlibrary{decorations.pathmorphing}
\usetikzlibrary{fit}
\usetikzlibrary{decorations.pathreplacing}

\newtoggle{quickdraw}
\toggletrue{quickdraw} 

\definecolor{lightgrey}{rgb}{0.9,0.9,0.9}
\definecolor{darkgreen}{rgb}{0,0.3,0}

\definecolorset{rgb}{}{}{darkred,0.8,0,0;darkgreen,0,0.5,0;darkblue,0,0,0.5}

\newcommand{\dx}{\mbox{d}}

\renewcommand{\vec}[1]{\mathbf{#1}}
\newcommand{\numTaxa}{N}
\newcommand{\numTraits}{D}
\newcommand{\numDatasets}{M}

\newcommand{\traitData}{\vec{Y}}

\newcommand{\datasetIndex}{m}
\newcommand{\exemplar}{\text{e}}

\newcommand{\sequences}{\vec{S}}
\newcommand{\latentData}{\vec{X}}
\newcommand{\latentdata}{\vec{x}}
\newcommand{\latentDatum}{x}
\newcommand{\phylogeneticParameters}{\boldsymbol{\phi}}
\newcommand{\phylogeny}{{\cal G}}
\newcommand{\tree}{\phylogeny}
\newcommand{\transpose}{^{t}}

\newcommand{\distanceMatrix}{\mathbf{Y}}
\newcommand{\distance}{y}
\newcommand{\summant}{r}

\newcommand{\cdensity}[2]{\ensuremath{p(#1 \,|\,#2)}}
\newcommand{\density}[1]{\ensuremath{p(#1 )}}

\newcommand{\treeNode}{\nu}

\newcommand{\traitVariance}{\mathbf{\Sigma}}
\newcommand{\nodeIndex}{c}

\newcommand{\rootMean}{\boldsymbol{\mu}_0}
\newcommand{\rootVarianceScalar}{\tau_0}
\newcommand{\unsequencedVarianceScalar}{\tau_{\exemplar}}
\newcommand{\treeVariance}{\vec{V}_{\tree}}

\newcommand{\mdsSD}{\sigma}
\newcommand{\mdsVariance}{\mdsSD^2}

\newcommand{\modelDistance}{\delta}

\newcommand{\normalCDF}[1]{\Phi \left( #1 \right)}

\newcommand{\order}[1]{{\cal O}\hspace{-0.2em}\left( #1 \right)}

\newcommand{\rootNode}{\nu^{\datasetIndex}_{2 \numTaxa_{\datasetIndex} -1 }}

\newcommand{\pathLengthNew}[2]{
	d_{F}
	(
	{#1}, {#2}
	)
}
\newcommand{\J}{\vec{J}}
\newcommand{\pprime}{^{\prime}}

\definecolor{trevorblue}{rgb}{0.330, 0.484, 0.828}
\definecolor{trevoryellow}{rgb}{0.829, 0.680, 0.306}

\title{Massive parallelization boosts big Bayesian multidimensional scaling}
\date{}

\author[1]{Andrew Holbrook}
\author[2]{Philippe Lemey}
\author[2]{Guy Baele}
\author[2]{Simon Dellicour}
\author[3]{Dirk Brockmann}
\author[4,5]{Andrew Rambaut}
\author[1,6,7]{Marc Suchard}
\affil[1]{Department of Human Genetics, University of California, Los Angeles}
\affil[2]{Department of Microbiology, Immunology and Transplantation, Rega Institute, KU Leuven}
\affil[3]{Institute for Theoretical Biology, Humboldt University Berlin}
\affil[4]{Institute of Evolutionary Biology, University of Edinburgh}
\affil[5]{Fogarty International Center,  National Institutes of Health}
\affil[6]{Department of Biostatistics, University of California, Los Angeles}
\affil[7]{Department of Biomathematics, University of California, Los Angeles}

\graphicspath{{figures/}}

\begin{document}
	
	\maketitle

	\clearpage
	
	\begin{abstract}
		Big Bayes is the computationally intensive co-application of big data and large, expressive Bayesian models for the analysis of complex phenomena in scientific inference and statistical learning.
		Standing as an example, Bayesian multidimensional scaling (MDS) can help scientists learn viral trajectories through space-time, but its computational burden prevents its wider use.
		Crucial MDS model calculations scale quadratically in the number of observations.
		We partially mitigate this limitation through massive parallelization using multi-core central processing units, instruction-level vectorization and graphics processing units (GPUs).
		Fitting the MDS model using Hamiltonian Monte Carlo, GPUs can deliver more than 100-fold speedups over serial calculations and thus extend Bayesian MDS to a big data setting.
		To illustrate, we employ Bayesian MDS to infer the rate at which different seasonal influenza virus subtypes use worldwide air traffic  to spread around the globe.
		We examine 5392 viral sequences and their associated 14 million pairwise distances arising from the number of commercial airline seats per year between viral sampling locations.
		To adjust for shared evolutionary history of the viruses, we implement a phylogenetic extension to the MDS model and learn that subtype H3N2 spreads most effectively, consistent with its epidemic success relative to other seasonal influenza subtypes.
		Finally, we provide \textsc{MassiveMDS}, an open-source, stand-alone \textsc{C++} library and rudimentary \textsc{R} package, and discuss program design and high-level implementation with an emphasis on important aspects of computing architecture that become relevant at scale.
	\end{abstract}
	
	\clearpage
	
	\section{Introduction}
	
The integral that so often arises from the application of Bayes' theorem exacerbates the general difficulties of big data statistical inference.
	Bayesian statisticians use the language of conditional probability to piece together large, flexible models with ease.
	Perversely, the integral and its myriad workarounds become less feasible as model complexity grows.
	Brute force computing rarely suffices for big Bayesian inference, the doubly dangerous collision between big data and massive model.
	As a field, Bayesian phylogeography illustrates  the challenges presented by big Bayes and the practical payoff in overcoming these challenges.
	
	The speed of transport in the global economy is matched by the complexity of travel patterns, which affect the emergence and spread of pathogens \citep{bloom2017emerging}.
	Scientists, epidemiologists, and policymakers must quickly visualize, draw actionable conclusions and make predictions from huge swaths of viral sequence data collected from all around the world.
	Although various approaches exist to study the spread of infectious diseases, recent developments for rapidly evolving pathogens take a probabilistic perspective simultaneously on the spatiotemporal spread and pathogen sequence mutation process.
	Here, phylogenetic diffusion models represent relatively simple and computationally efficient yet flexible tools to connect spatial dynamics to sequence evolution, and specific implementations of random walk models are available for both discrete and continuous location data for the sampled sequences \citep{lemey09,lemey2010phylogeography}.
	
	The discrete approach models transitioning between a limited set of discrete states throughout the ancestral history of the pathogen sequences according to a continuous-time Markov chain process and has facilitated the study of global movement patterns of influenza viruses, with most efforts for human flu focusing on influenza A/H3N2 \citep{bahl2011temporally,nelson2015global}.
	A recent study has broadened the focus to all four seasonal influenza viruses, including two influenza A subtypes (H3N2 and H1N1) and two influenza B subtypes (Yamagata and Victoria) \citep{bedford2015global}.
	This study shows the viruses varying in their degree of persistence and discretized-location switching frequency, with A/H1N1 evincing lower switching rates compared to A/H3N2 and the two B subtypes showing even lower rates.
	The study associates these differences with  how quickly the viruses evolve antigenically.
	A key element in the connection between antigenic drift and global movement are differences in age distributions of infection and  age-specific mobility patterns: viruses capable of evolving faster antigenically will be better at infecting adults, who tend to travel more frequently than children, providing more opportunities for the virus to spread \citep{bedford2015global}.
	
	Discrete phylogeographic reconstructions have important limitations such as their sensitivity to sampling biases and the need to specify arbitrary spatial partitions.
	The \emph{continuous} diffusion model offers an interesting alternative in this respect, but geographic space is ill-suited for tracking pathogens in humans and other hosts that frequently travel long distances.
	However, a recent modeling study has demonstrated that complex spatiotemporal patterns of spread evince surprisingly simple and regular wave-like patterns for distances measured along transport networks instead of spatial distances \citep{brockmann2013hidden}.
	
	We use this concept of `effective distance' to perform phylogeographic inference in latent effective space by adopting a Bayesian MDS approach that enables the quantification and comparison of differences in the rate at which the different seasonal influenza variants travel the air transportation network.
	Bayesian MDS  \citep{oh2001bayesian} probabilistically projects relationships between high dimensional objects onto low-dimensional space and thus accounts for uncertainty by integrating over constellations of latent locations.
	To adjust for shared evolutionary history between viral samples, \cite{bedford2014integrating} combines Bayesian MDS with a latent phylogenetic diffusion model.
	With pairwise distances between flu samples arising from biochemical assays, those authors used their Bayesian phylogenetic MDS model to draw insights into the changing antigenicity of the four major seasonal influenza subtypes.
	
	Unfortunately, the Markov chain Monte Carlo (MCMC) computations required for a Bayesian MDS (phylogenetic or otherwise) analysis are onerous.
	The MDS likelihood prevents Gibbs sampling, and complexity of the likelihood evaluations necessary for Metropolis-Hastings grows quadratically with the data.  \citet{fosdick2019multiresolution} circumvent MCMC computations by fitting the MDS model with an optimization routine.
	\cite{bedford2014integrating} partly avoid the issue by completely leaving out the truncation term on the non-negative pairwise distances---the computational bottleneck of the likelihood evaluation---and effectively draw inference based on an incorrect model.
	Those authors witness high autocorrelation between Markov chain states regardless.
	Hamiltonian Monte Carlo (HMC) \citep{neal2011mcmc}, an advanced MCMC algorithm that uses gradient information to craft proposals, could help improve this poor mixing, but the log-likelihood gradient evaluations required by HMC also scale quadratically with the data.
	
	Nonetheless, we assert that correct inference from Bayesian MDS is possible, even scalable, with the help of massively parallel computing that we exploit here to quickly calculate the MDS likelihood and log-likelihood gradient in the context of HMC.
	We are not the first to investigate parallel implementations in statistical computing: \cite{suchard2009many}, \cite{suchard2010understanding} and \cite{suchard2010some} apply graphics processing unit (GPU) computing to optimization and Bayesian inference for phylogenetics and flow cytometry; \cite{lee2010utility} perform sequential Monte Carlo with the aid of GPUs; \cite{zhou2010graphics} leverage GPUs for statistical optimization; and \cite{beam2016fast} compute the likelihood and its gradient for a multinomial model with GPUs and thus accelerate HMC for that model.
	
	While GPUs also deliver the greatest speed gains for the inference problem considered here, we find that multi-core central processing units (CPUs) combined with on-chip vectorization follow closely behind in scaling Bayesian MDS for millions of data points.
	To facilitate adoption of both GPU- and CPU-based parallel computing for Bayesian MDS, we provide the open-source library \textsc{MassiveMDS} \url{http://github.com/suchard-group/MassiveMDS} both as an \textsc{R} package and as a stand-alone \textsc{C++} library.
		Section \ref{sec:software} contains further information on \textsc{MassiveMDS} and related software that we provide readily available online. We now introduce Bayesian MDS and its phylogenetic instantiation.
	
	\section{Methods}

	\subsection{From dissimilarity to a latent space}
	
	Multidimensional scaling (MDS) encompasses a class of ordination methods that project a collection of objects into a low-dimensional Euclidean space based on dissimilarity measurements between pairs of objects \citep{kruskal1964multidimensional}.
	Through MDS, objects with smaller dissimilarity generally find themselves nearer in $L_2$ distance to each other in this latent space than objects with larger dissimilarity.
	While MDS traditionally has found use as an exploratory data analysis tools, model-based MDS variants exist within the Bayesian framework \citep{desarbo1998bayesian,oh2001bayesian}.
	One posits that each object's latent location is a random variable, translates the MDS projection into a probability model on the observed dissimilarities given distances between latent locations \citep{ramsay1982some} and specifies an appropriate prior distribution over these locations.
	Previously, such distributions have remained arbitrary and relatively uninformative.
	In this paper, however, the stochastic process that gives rise to the latent location prior distribution is highly-structured and well-informed.
	Further, the parameters that characterize the process are of chief scientific interest.
	
	We are interested in a finite collection of $N$ items. For any two distinct items $i$ and $j$, we follow \citet{oh2001bayesian} and model the observed dissimilarity $\distance_{ij}$ as conditionally independent, normal random variables, truncated to be positive
	\begin{align}
	\distance_{ij} \sim \mbox{N} \left( \modelDistance_{ij}, \mdsVariance \right) \mbox{I} \left( \distance_{ij}  > 0 \right) \text{ for } i > j,
	\end{align}
	where the expected dissimilarity $\modelDistance_{ij} = || \latentdata_i - \latentdata_j ||$ is the $L_2$ norm between latent locations $\latentdata_i = (\latentDatum_{i1}, \ldots, \latentDatum_{i \numTraits} )\transpose$  and $\latentdata_j  = (\latentDatum_{j 1}, \ldots, \latentDatum_{j \numTraits} )\transpose$ 
	in a low-dimensional, real coordinate space $\mathbb{R}^{\numTraits}$.
	Given all latent locations $\latentData = \left( \latentdata_{1}, \ldots, \latentdata_{\numTaxa} \right)\transpose$, the conditional density of the observed data $\distanceMatrix$ becomes
	\begin{align}
	\cdensity{\traitData}{\latentData, \mdsVariance} & \propto
	\left(
	\mdsVariance
	\right)
	^{\frac{\numTaxa ( 1-\numTaxa) }{4}}
	\exp
	%
	%
	\left(	-
	\sum_{i > j}
	\summant_{ij}
	\right)
	\nonumber \\
	\summant_{ij} &=
	\frac{ \left( \distance_{ij} - \modelDistance_{ij} \right)^2 }{ 2 \mdsVariance }
	+ \log  \normalCDF{ \frac{\modelDistance_{ij}}{ \mdsSD} }
	,
	\label{eq:likelihood}
	\end{align}
	where $\normalCDF{\cdot}$ is the cumulative distribution function of a standard normal random variable.
	
	Motivation behind this probabilistic transformation from observed dissimilarity into a Euclidean space rests on present limitations in drawing inference about diffusive processes (such as those outlined below) over irregular landscapes \citep{billera01}.
	Inference over irregular landscapes often necessitates extensive data augmentation or numerical integration \citep{manton2013primer,nye2014diffusion} to closely approximate the density function of a partially observed sample-path, as the density function is a solution of the stochastic differential equation that governs the diffusion.
	On the other hand, in a latent Euclidean space, simple Brownian diffusion \citep{brown1828,wiener58} and its scale mixtures generalization \citep{lemey2010phylogeography} offer closed-form density functions.  Further, these functions yield conveniently to the analytic integration necessary to track the multiple end-state locations $\latentData$ of dependent diffusion processes.
	
	\subsection{Highly-structured Brownian process prior}
	
	Consider a collection of $\numDatasets$ molecular sequence alignments, where each alignment $\sequences_{\datasetIndex}$ for $\datasetIndex = 1,\ldots,\numDatasets$ contains sequences from $\numTaxa_{\datasetIndex}$ evolutionarily related viruses, $\numTaxa_{\exemplar}$ additional unsequenced viruses from unsampled locations and an $\numTaxa \times \numTaxa$ symmetric, dissimilarity matrix $\distanceMatrix$ for $\numTaxa = \sum_{\datasetIndex=1}^{\numDatasets} \numTaxa_{\datasetIndex} + \numTaxa_{\exemplar}$, where
	in general $\numTaxa \gg \numTaxa_{\exemplar}$.
	Each entry $\distance_{ij}$ of $\distanceMatrix$ reports a non-negative dissimilarity measurement between virus $i$ and $j$ for $i,j = 1,\ldots, N$.
	\newcommand{\whichPhylogeny}[1]{\phylogeny\left( #1 \right)}
	We follow standard Bayesian phylogenetics hierarchical approaches \citep{Suchard03HPM} to model the sequence data $\sequences = \left(\sequences_1, \ldots, \sequences_{\numDatasets} \right)$ that include, among other parameters $\phylogeneticParameters$ not critical to the development in this paper, phylogenetic trees $\phylogeny = (\phylogeny_1, \ldots, \phylogeny_{\numDatasets} )$ which may be known \textit{a priori} or random.
	Each tree $\phylogeny_{\datasetIndex}$ is a bifurcating, directed graph with $N_{\datasetIndex}$ terminal degree-1 nodes $(\treeNode^{\datasetIndex}_{1}, \ldots, \treeNode^{\datasetIndex}_{N_{\datasetIndex}})$ that correspond to the tips of the tree, $N_{\datasetIndex} - 2$ internal degree-3 nodes $(\treeNode^{\datasetIndex}_{N_{\datasetIndex}+1}, \ldots, \treeNode^{\datasetIndex}_{2N_{\datasetIndex}-2})$, a root degree-2 node $\treeNode^{\datasetIndex}_{2N_{\datasetIndex}-1}$ and edge weights $(t^{\datasetIndex}_{1},\ldots, t^{\datasetIndex}_{2N_{\datasetIndex}-2})$ that report the elapsed evolutionary time between nodes.
	To simplify notation later, let $\whichPhylogeny{i}\, \in\, \{1,\ldots,\numDatasets\}$ indicate the alignment to which virus $i$ belongs, with $0$ indicating $i$ unsampled.
	We assume conditional independence between $\sequences$ and $\distanceMatrix$ given $\phylogeny$.
	Interested readers may explore, for example, \citet{suchard01} or \cite{suchard2018bayesian} for detailed development of $\density{\sequences, \phylogeneticParameters, \phylogeny}$.
	
	\begin{figure}[t]
		\begin{center}
			\includegraphics[width=\textwidth]{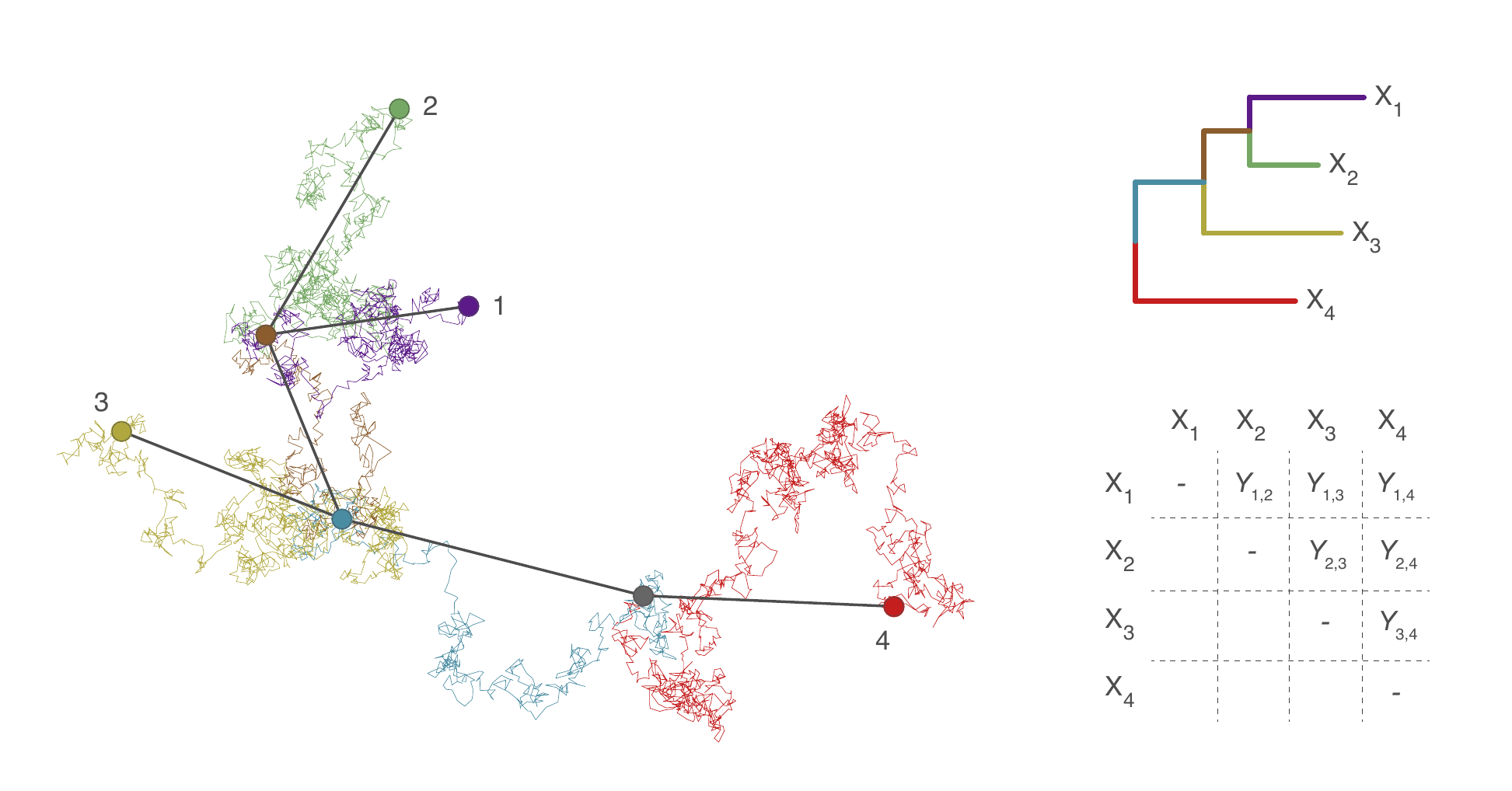}
		\end{center}
		\caption{Multivariate Brownian diffusion along a phylogeny as a latent Gaussian model prior.
			This example phylogeny has 4 tips, labeled 1, 2, 3 and 4.
			We depict the conditionally independent diffusion realizations in two dimensions along each branch in different colors.
			The root and two internal node realizations are colored as gray, blue and brown circles, while the four tip node realizations $\latentData = \left(\latentdata_{1}, \latentdata_{2}, \latentdata_{3}, \latentdata_{4}\right)$ are highlighted in purple, green, yellow and red, respectively.
			Dynamic programming enables us to integrate over all possible root and internal node realizations, returning the joint distribution of $\latentData$ as an informed prior.
			Pairwise distances between tip realizations relate to the observed dissimilarity distances $\distance_{ij}$ of the model.
		}
		\label{fig:gp}
	\end{figure}
	
	Our model posits that a multivariate Brownian diffusion process along the branches of the trees in $\phylogeny$ \citep{lemey2010phylogeography} gives rise to $\latentData$.
	The Brownian process asserts that the latent location value of a child node $\treeNode^{\datasetIndex}_{\nodeIndex}$ in tree $\phylogeny_{\datasetIndex}$ is multivariate normally distributed about the latent value of its parent node $\treeNode^{\datasetIndex}_{\text{\tiny pa}(\nodeIndex)}$ with variance $t^{\datasetIndex}_{\nodeIndex} \times \traitVariance$.
	The unknown $K \times K$ matrix $\traitVariance$ parameterizes the dispersal rate in the latent space after controlling for correlation in latent values that are shared by descent through $\phylogeny_{\datasetIndex}$.
	This construction generalizes univariate Comparative Methods \citep{cavalli1967phylogenetic,felsenstein85} approaches to model the evolution of continuous-valued random variables first into a multivariate setting and second across multiple trees.
	Figure \ref{fig:gp} illustrates one possible realization of this process for a single tree with $\numTaxa_{\datasetIndex} = 4$ tips.
	
	We assume that the latent values at all $\numDatasets$ root nodes $\treeNode^{\datasetIndex}_{2N_{\datasetIndex}-1}$ and for the $\numTaxa_{\exemplar}$ unsequenced viruses are \textit{a priori} multivariate normally distributed with mean $\rootMean$ and variance $\rootVarianceScalar \times \traitVariance$ or $\unsequencedVarianceScalar \times \traitVariance$, respectively. 
	Following \citet{cybis2015assessing}, we can ascribe that jointly $\latentData$ is matrix normally distributed, with probability density function
	\newcommand{\sameTree}[2]{\delta_{#1 #2}}
	\begin{align} \small
	\cdensity{\latentData}{ \treeVariance, \traitVariance, \rootMean, \rootVarianceScalar, \unsequencedVarianceScalar}
	=
	\frac{
		\mbox{exp}
		\left\{
		-\frac{1}{2} \mbox{tr}
		\left[
		\traitVariance^{-1}
		\left( \latentData - \rootMean \right)^{t}
		\treeVariance^{-1}
		\left( \latentData - \rootMean \right)
		\right]
		\right\}
	}{
		\left(2 \pi\right)^{\numTaxa\numTraits/2}
		\left| \traitVariance \right| ^{\numTaxa/2}
		\left|
		\treeVariance
		\right|^{\numTraits/2}
	} ,
	\label{eq:multinormal}
	\end{align}
	where
	$\treeVariance = \{ v_{i j} \}$ is a block-diagonal $\numTaxa \times \numTaxa$ matrix. 
	Specifically, if virus $i$ is unsequenced, then $v_{ii} = \unsequencedVarianceScalar$ and $v_{ij} = 0$ for all $j \neq i$.
	The trees in $\tree$ define the remaining $\numDatasets$ blocks.
	We define $\pathLengthNew{u}{w}$ to equal the edge-weight sum along the shortest path between node $u$ and node $w$ in a tree $\tree_{\datasetIndex}$.
	%
	Then, within block $\datasetIndex$,
	diagonal elements $v_{i\pprime i\pprime} = \rootVarianceScalar + \pathLengthNew{\rootNode}{\nu_{i\pprime}}$, the elapsed evolutionary time between the root node and tip node $i\pprime$, and off-diagonal elements $v_{i\pprime j\pprime} = \rootVarianceScalar +
	\left[
	\pathLengthNew{\rootNode}{\nu_{i\pprime}} + \pathLengthNew{\rootNode}{\nu_{j\pprime}}
	- \pathLengthNew{\nu_{i\pprime}}{\nu_{j\pprime}}
	\right] / 2$, the elapsed time between the root 
	and the most recent common ancestor of tip nodes $i\pprime$ and $j\pprime$.

	To complete our model specification, we assume \textit{a priori}
	\begin{align}
	\traitVariance^{-1} & \sim \mbox{Wishart}(d_0, \vec{T}_0) \text{ and} \nonumber \\
	\mdsSD^{-2} & \sim \mbox{Gamma}(s_0, r_0),
	\end{align}
	with degrees of freedom $d_0$, rate matrix $\vec{T}_0$, shape $s_0$ and rate $r_0$.
	Finally, we specify fixed hyperparameters $(\rootMean, \rootVarianceScalar, d_0, \vec{T}_0, s_0, r_0)$ in our example.

	\subsection{Inference}
	
	We use Markov chain Monte Carlo (MCMC) to learn the posterior distribution
	\begin{align}
	\label{eq:posterior}
	\cdensity{\traitVariance, \mdsVariance, \tree, \phylogeneticParameters}{\traitData, \sequences}
	& \propto
	\cdensity{\traitData}{\traitVariance, \mdsVariance, \tree}
	\times \density{\traitVariance} \times \density{\mdsVariance}
	\times \density{\sequences, \phylogeneticParameters, \tree}
	\\ \nonumber
	& =
	\left(
	\int
	\cdensity{\traitData}{\latentData, \mdsVariance}
	\cdensity{\latentData}{\traitVariance, \tree}
	\dx \latentData
	\right)
	\times \density{\traitVariance} \times \density{\mdsVariance}
	\times \density{\sequences, \phylogeneticParameters, \tree}
	\end{align}
	with a random-scan Metropolis-with-Gibbs scheme and the development of a computationally efficient transition kernel to sample the latent values $\latentData$.
	We exploit standard Bayesian phylogenetic algorithms \citep{suchard2018bayesian} based on Metropolis-Hastings sampling for the tree $\tree$ and other phylogenetic parameters $\phylogeneticParameters$.
	These latter transition kernels are not rate-limiting.
	
	\par
	
	Sampling $\tree$ necessitates evaluating $\cdensity{\latentData}{\traitVariance, \tree}$.
	Equation (\ref{eq:multinormal}) suggests a computational order $\order{ \tilde{\numTaxa}^3 }$ where $\tilde{\numTaxa} = \max\limits_{\datasetIndex} \numTaxa_{\datasetIndex}$
	to form the matrix inverse $\treeVariance^{-1}$.
	However, we follow \citet{pybus2012unifying} who develop a dynamic programming algorithm to evaluate Equation (\ref{eq:multinormal}) in $\order{ \tilde{\numTaxa} \numTraits^2 }$ via parallelizable post-order traversals of the trees in $\tree$.
	\citet{freckleton2012fast} and \citet{ho2014linear} propose similar linear-time algorithms, but the underlying idea of message passing on a directed, acyclic graph extends back at least to \citet{cavalli1967phylogenetic} and \citet{pearl1982reverend}.
	Sampling $\phylogeneticParameters$ is, likewise, linear in $\tilde{\numTaxa}$ and also conveniently computable on massively parallel devices \citep{suchard2009many}.
	
	On the other hand, sampling $\latentData$ stands as the rate-limiting operation for posterior inference.
	To appreciate why, the full conditional distribution $\cdensity{ \latentData }{ \traitData, \tree, \traitVariance, \mdsVariance}$ is not of standard form, eliminating Gibbs sampling.
	Any sampler for $\latentData$ must therefore evaluate or approximate $\log \cdensity{\traitData}{\latentData, \mdsVariance}$
	or its derivatives.
	This log-density is a sum over $\numTaxa (\numTaxa - 1) / 2 $ terms, where each involves $\order{\numTraits}$ operations, such that complete evaluation is $\order{\numTaxa^2 \numTraits}$.  Its gradient is $\numTaxa$ summations over $\numTaxa-1$ terms that also involve $\order{\numTraits}$ operations. Again, complete evaluation is $\order{\numTaxa^2 \numTraits}$.

	\citet{bedford2014integrating} describe a previous attempt at fitting a related Bayesian MDS model with a phylogenetically informed prior.
	Restricted to a modest $\numTaxa \sim \text{low }100\text{s}$ similar to the problems that \citet{oh2001bayesian} attack, this previous work has relied on a low-dimensional random-walk transition kernel applied at random to a single element $\latentdata_i$ or even $x_{i k}$ of $\latentData$.
	The advantage here exploits caching the $\order{\numTaxa^2}$ terms, since changing the value of a single $\latentdata_i$ invalidates only $\numTaxa - 1$ terms.
	The disadvantage lies in the potentially extreme auto-correlation of the MCMC sample that grows with increasing $\numTaxa$.
	The auto-correlation arises out of the posterior dependence between $\latentdata_i$ and $\latentdata_j$ for $i \neq j$ that our highly-informative phylogenetic prior $\cdensity{\latentData}{\traitVariance, \tree}$
	exacerbates.

	\paragraph{Hamiltonian Monte Carlo}
	
	\par
	
	\newcommand{\targetDistribution}[1]{\pi \left( #1 \right)}
	\newcommand{\nextDraw}{\boldsymbol{\nu}}
	\newcommand{\genericRV}{\latentdata}
	\newcommand{\momentum}{\mathbf{p}}
	\newcommand{\mass}{\mathbf{M}}
	
	We rely on HMC \citep{duane1987hybrid,neal2011mcmc}  to efficiently simulate from the posterior distribution with respect to latent data $\latentData$.  HMC is an advanced MCMC methodology that employs deterministic Hamiltonian dynamics to intelligently generate proposal states.  We couple each trajectory with a Metropolis accept-reject step that renders the target distribution invariant.  The upshot is an algorithm that has aided Bayesian analysis by facilitating posterior inference for models of unprecedented dimension and hierarchical structure.
	
	Specifically, let $\pi(\genericRV)$ be the probability density function of the target distribution for generic random variable $\genericRV$ with continuous support.  Further, assume that $\pi(\cdot)$ be differentiable, as is the case for the application considered here.  HMC works by augmenting the `position' variable $\genericRV$ by an independent, auxiliary `momentum' variable $\momentum$ with Gaussian density  $\xi(\momentum)$.  An energy function is constructed as the negative logarithm of the density for the joint distribution over $(\genericRV,\momentum)$:
	\begin{align}
	H(\genericRV,\momentum)= - \log \big(\pi(\genericRV)\, \xi(\momentum) \big) \propto -\log \pi(\genericRV) + \frac{1}{2} \momentum^T\mass^{-1}\momentum \, .
	\end{align}
	Here, $\mathbf{M}$ is the covariance of $\momentum$, but it is also interpretable as the mass matrix for the dynamical system associated to Hamiltonian $H(\genericRV,\momentum)$ and described by the system of equations
	\begin{align}
	\dot{\genericRV} &= \frac{\partial}{\partial \momentum}H(\genericRV,\momentum) = \frac{1}{2} \mass^{-1} \momentum \\ \nonumber
	\dot{\momentum} &= - \frac{\partial}{\partial \genericRV}H(\genericRV,\momentum) = \nabla \log \pi(\genericRV) \, .
	\end{align}
	By Liouville's theorem, Hamiltonian dynamics conserve the energy $H(\genericRV,\momentum)$. It is a corollary that perfect simulation of the Hamiltonian system is equivalent to perfect sampling from the canonical distribution $\exp \left(-H(\genericRV,\momentum) \right) = \pi(\genericRV) \, \xi(\momentum) $, i.e., for any stopping time $t$, the Metropolis-Hastings acceptance criterion is
	\begin{align}
	\min \left(1, \,\frac{\exp \left(-H(\genericRV_t,\momentum_t)\right)}{\exp \left(-H(\genericRV_0,\momentum_0)\right)} \right) = \min (1,\,1) = 1\, .
	\end{align}
	Exact simulation is rarely available, but the discretized leapfrog algorithm \citep{leimkuhler2004simulating} has proven effective for simulating the Hamiltonian dynamics.  The discretization \emph{does} lead to errors which cause the Hamiltonian evaluated at the proposed state $(\genericRV^*,\momentum^*)$ to differ from that of the current state, so an accept-reject step remains necessary.  On the other hand, the discretized dynamics preserve volume just like the true Hamiltonian dynamics, so we need no Jacobian corrections to  calculate the Metropolis acceptance criterion.  We do, however, need to evaluate the log-likelihood gradient at multiple points along the trajectory.
	
	\subparagraph{The log-likelihood gradient}\label{ap:gradient}
	\newcommand{\gradContribution}{\mathbf{r}_{ij}}
	
	In the notation for Bayesian MDS, HMC requires the gradient of the log-likelihood with respect to latent locations $\latentData$. The gradient of the log-likelihood with respect to a single row $\latentdata_i$ of the matrix $\latentData$ is
	\begin{align}\label{eq:gradient}
	\frac{\partial}{\partial \latentdata_i} \log \cdensity{\traitData}{\latentData, \mdsVariance} &= \frac{\partial}{\partial \delta_{ij}} \log \cdensity{\traitData}{\latentData, \mdsVariance} \, \frac{\partial \delta_{ij}}{\partial \latentdata_i}  \\ \nonumber
	&= - \sum_{j \neq i} \left( \frac{(\delta_{ij}-\distance_{ij})}{\mdsVariance} +\frac{\phi(\delta_{ij}/\sigma)}{\sigma \Phi(\delta_{ij}/\sigma)} \right) \frac{\partial \delta_{ij}}{\partial \latentdata_i}  \\ \nonumber
	&= - \sum_{j \neq i} \left( \frac{(\delta_{ij}-\distance_{ij})}{\mdsVariance} +\frac{\phi(\delta_{ij}/\sigma)}{\sigma \Phi(\delta_{ij}/\sigma)} \right) \frac{(\latentdata_i-\latentdata_j)}{ \delta_{ij}}  \\ \nonumber
	&:= - \sum_{j \neq i} \gradContribution \, .
	\end{align}
	Here, $\phi(\cdot)$ is the probability density function of a standard normal variate, and $\gradContribution$ is the contribution of the $j$th location to the gradient with respect to the $i$th location.

	\subsection{Model selection}
	
	\newcommand{\otherParams}{\Theta}
	
	We use cross-validation when it is difficult \emph{a priori} to motivate a modeling decision.  In particular, we are interested in judging the `correct' number of dimensions available for the latent diffusion process.  In turn, we interpret this latent dimensionality as a rough quantification of the complexity for the air-traffic space in junction with the pathogen's evolutionary dynamics.
	
	In cross-validation \citep{geisser1975predictive}, the practitioner systematically excludes a fixed proportion of observations, trains the model on those remaining observations and uses the fitted model to predict the held-out data.  We repeat this process over a number of `folds' and ascertain the prediction error over the different folds.  The model with the smallest total prediction error is deemed best.
	
	Recall that, for MDS, our data is a large distance matrix $\mathbf{Y}$, and observations correspond to off-diagonal elements $\distance_{ij}$.  A cross-validation fold consists of the held-out observations $\distanceMatrix^f_{IJ}$ and the remaining observations  $\distanceMatrix^f_{-(IJ)}$ for some multi-index $IJ$ depending on $f$.  Let $s$ index an MCMC state corresponding to a single draw from the target posterior and denote the complete set of latent locations and model parameters $(\latentData,\otherParams)^f_s$ for $s=1,\dots,S^f$.  We use the empirical \emph{log pointwise predictive density} ($\widehat{lpd}$) as measure of predictive accuracy and model fit \citep{vehtari2017practical}.   The log pointwise predictive density is
	\begin{align*}
	lpd &= \sum_f \sum_{\substack{ i<j \\ i,j \in IJ}} \log p(\distance_{ij}^f|\distanceMatrix_{-(IJ)}^f)  \\
	&= \sum_f  \sum_{\substack{ i<j \\ i,j \in IJ}} \log \int p(\distance^f_{ij}|(\latentData,\Theta)) p((\latentData,\Theta)|\distanceMatrix_{-(IJ)}^f) \, \dx (\latentData,\Theta) \\
	&\approx \sum_f  \sum_{\substack{ i<j \\ i,j \in IJ}} \log \frac{1}{S^f} \sum_{s=1}^{S^f} p(\distance^f_{ij}|(\latentData,\Theta)^f_s) p((\latentData,\Theta)^f_s|\distanceMatrix^f_{-(IJ)}) = \widehat{lpd} \, .
	\end{align*}
	Given two models with differing latent dimensions, we choose the model with smaller $\widehat{lpd}$.  For more on Bayesian model selection, see \cite{gelman2013bayesian}.

	\newcommand{\transformR}{r}
	\newcommand{\transformCDF}{c}
	\newcommand{\threadsPerBlock}{B}
	
	\subsection{Massive parallelization}
	\label{sec:parallelization}

	\algblock{ParFor}{EndParFor}
	\algnewcommand\algorithmicparfor{\textbf{parfor}}
	\algnewcommand\algorithmicpardo{\textbf{do}}
	\algnewcommand\algorithmicendparfor{\textbf{end\ parfor}}
	\algrenewtext{ParFor}[1]{\algorithmicparfor\ #1\ \algorithmicpardo}
	\algrenewtext{EndParFor}{\algorithmicendparfor}

	HMC is a powerful tool for Bayesian learning, but the likelihood and gradient evaluations it necessitates become overly burdensome for Big Data inference.
	Efficiently computing $\log \cdensity{\traitData}{\latentData, \mdsVariance}$ and its gradient with respect to $\latentData$ remains a critical and rate-limiting step at $\order{\numTaxa^2 \numTraits}$ for each evaluation.  When $\numTaxa$ is large, the key insights for effective parallelization of these evaluations are three-fold.
	
	The first is most important: for the likelihood, there are a massive number of stereotyped, and seemingly independent, operations in evaluating $\summant_{ij}$ for all $i > j$ (a transformation); for the gradient, there are a massive number of stereotyped, and seemingly independent, operations in evaluating $\gradContribution$ for all $i > j$ (again, a transformation).  For both $\summant_{ij}$ and $\gradContribution$, the floating point operations required to evaluate $\Phi(\cdot)$ are rate-limiting, so a useful computing strategy limits wall time spent performing these operations by applying the functions in parallel across a range of inputs.
	
The final two insights are often missed in statistical computing.
		The above transformations offer a high degree of data-reuse; for example, $\summant_{ij}$ and $\gradContribution$ for all $j$ depend on $\latentdata_i$ and $\distance_{ij}$, and $\summant_{ij}$ and $\gradContribution$ for all $i$ depend on $\latentdata_j$ and $\distance_{ij}$.  A good computing strategy stores these values in a way that facilitates fast  reuse.
		Finally, the realized values of $\summant_{ij}$ and  $\gradContribution$ are never actually needed, only their sum (a reduction).  A good strategy avoids storing intermediate values in costly memory.

	\paragraph{Parallelization strategies}
	
	The simultaneous execution of multiple mathematical operations through parallelization continues to enable computation to keep pace with Moore's Law that posits processing power doubles approximately every two years.
	Parallelization is growing into a dominant theme in large-scale statistical inference \citep{suchard2010some} and on hardware including clusters of independent compute nodes, on-chip vector instructions, multithreaded multi-core processors and parallel co-processors such as GPUs.
	Capitalizing on these hardware features in software implementation could emerge as the most important task facing a computational statistician.
	
	Cluster computing is the most familiar form of parallelization and often scales up to 1000s or more of nodes that contain their own CPUs and random access memory (RAM), linked loosely together through an Ethernet or InfiniBand network.
	In total, 1000s of CPUs and massive quantities of RAM can comprise a single cluster; however, sharing information between individual nodes has high latency in these distributed-computing environments.
	Communication latency can severely affect the theoretically achievable speed-up of parallelization.
	
	\newcommand{\numThreads}{S}
	\newcommand{\thread}{s}
	
	Assume a computing task has an overall execution cost $c_0$ and can be evenly distributed across $\numThreads$ parallel devices, each executing a thread of work.
	Then, an ideal implementation will still require actual time $c_0 / \numThreads + c_1$, where $c_1$ is the incurred time of communication or additional non-parallelizable computation.
	The speed-up on this ideal system often falls far short of $\numThreads$-fold.
	Further, the overhead parameter $c_1$ often scales with $\numThreads$.
	The high latency of clusters, therefore, suggests coarse-grain decomposition with larger and relatively independent threads assigned to each compute node to minimize the parallelization overhead.
	Further, many statistical model fitting algorithms, including high-dimensional optimization and MCMC, are iterative such that $c_1$ also scales by the number of iterations.
	While several tools stand out for cluster-based statistical computing \citep{schmidberger2009state}, we often find that significant financial investment in purchasing or renting a large cluster yields only modest speed-up.
	For example, \citet{suchard2010understanding} benchmark a related MCMC inference problem involving millions of observations, yielding an approximate 20-fold speed-up on a 100-node cluster costing about \$250,000 in 2008, or about \$3,600 for a month of compute-time today on a cloud-based service.
	
	In this paper, we avoid the high latency cost of iterative algorithms on clusters and instead raise awareness on exploiting parallelization in less expensive shared-computing environments.
	Specifically, we explore two prominent and one often unrecognized avenues of parallelization to evaluate $\log \cdensity{\traitData}{\latentData, \mdsVariance}$ (Equation  \ref{eq:likelihood}) and its gradient with respect to $\latentData$ (Equation \ref{eq:gradient}), our rate-limiting steps in inference.
	On the prominent side, these avenues harness the multiple, multithreaded processing cores cast onto CPUs in standard desktop computers and the extreme number of cores working in tandem on GPU cards now ubiquitous as high-performance computing add-ons.
	Under appreciated stands single-instruction, multiple-data (SIMD) integer or floating-point operation parallelization available within standard CPU cores.
	
	In the parlance of high-performance computing, evaluating Equation (\ref{eq:likelihood}) is a transformation-reduction, where the transformation reads from RAM the latent locations $\latentData$ and observed distances $\distanceMatrix$ and computes $\summant_{ij}$ for all $i > j$ and the reduction then sums together all $\summant_{ij}$. A first glance at this computation suggests two separate loops over all $i > j$; the first presents $\order{\numTaxa^2}$ embarrassingly parallel tasks with floating point math calls, and the second loop has strong serial dependence.
	Gradient evaluation requires the $\numTaxa$  transformation-reductions given by Equation \eqref{eq:gradient} for each $i$. Again, each transformation-reduction suggests two separate loops over all $i > j$; here, the inner loop is $\order{\numTaxa}$ with floating point math operations, and the outer loop again has strong serial dependence.
	
	Rate-limiting are the embarrassingly parallel floating point operations characterized by math function calls.  On account of their heavy cost, it is advantageous to broadcast such expensive operations across as broad a swath of input data as possible. GPUs efficiently apply operations to thousands of inputs, and, with the help of SIMD, multi-core CPUs simultaneously apply a single floating operation to independent inputs in the low hundreds.   Parallel computing of floating point operations is much faster than serial, but an important caveat of high-performance computing is that a single memory transaction involving a read or write from RAM may take up to two orders-of-magnitude more time than a numerical operation applied to a value sitting in a limited number of storage locations called registers within the CPU or GPU.
	In practice, multiple consecutive memory transactions occur simultaneously, moving a block of data between RAM and successive layers of high-speed memory in close proximity to the processor called caches.
	As a result, consecutive memory transactions are considerably faster than random ones, but still pale in speed comparison to on-chip numerical operations.
	
	To avoid unnecessary memory transactions, it should become common practice in statistical computing to ``fuse'' together performance-dependent transformation-reductions into single loops even if they retain some serial dependence.
	Through fusion, for example, intermediate $\summant_{ij}$ values never need to be written to RAM in the transformation, nor read back for the reduction.
	Instead, one or more partials sums remain in on-chip registers and get incremented as intermediate values become available; these may become available in a parallel fashion and, ultimately, we may reduce either the intermediate values or final partial-sums in parallel over a binary tree. 
	Given $B$ intermediates or partial-sums, the computational order of the binary-tree reduction is $\order{\log B}$, but it carries high inter-thread communication.
	On hardware with even modest communication latency between threads, such as between multiple cores on CPUs, there is often little speed advantage to a binary-tree reduction.
	On the other hand, with almost no communication latency between small groups of threads on GPUs, the final parallel reduction often shines.
	We return to this point below.
	
	\paragraph{Multi-core CPUs}
	
	\begin{figure}[t]
		\centering
		\begin{tikzpicture}[
		myrect/.style={
			rectangle,
			draw,
			inner sep=0pt,
			fit=#1}
		]
		\pgfmathsetseed{666}
		\def\maxrand{99}
		
		\tikzstyle{Sample} = [
		draw, anchor=west,
		inner sep=0,
		outer sep=0,
		minimum height=\sampleheight * 1cm,
		font=\small,
		text=black,
		]
		
		\pgfmathsetmacro{\totalwidth}{10}
		\pgfmathsetmacro{\levelheight}{2.4}
		
		\pgfmathsetmacro{\sampleheight}{1.0}
		\pgfmathsetmacro{\samplewidth}{0.5}

		\definecolor{lowcolor} {rgb}{0.6,0.6,1}
		\definecolor{highcolor}{rgb}{0.6,1,0.6}
		
		\definecolor{outLowcolor} {rgb}{1,1,0.8} 
		\definecolor{outHighcolor}{rgb}{0.6,0.2,0.02}	
		
		\def\globalsize{64}
		\pgfmathsetmacro{\sampleheight}{10.0 / \globalsize * 2.0}
		\pgfmathsetmacro{\samplewidth}{10.0 / \globalsize}
		
		\pgfmathsetmacro{\outputheight}{6.0 / \globalsize}
		\pgfmathsetmacro{\outputwidth}{6.0 / \globalsize}
		
		\pgfmathtruncatemacro{\runningrandarray}{random(\maxrand)}
		\foreach \x in {1,...,\globalsize}  {
			\let\temprand\runningrandarray
			\pgfmathtruncatemacro{\tempres}{random(\maxrand)}
			\xdef\runningrandarray{\temprand,\tempres}
		}
		\xdef\randarray{{\runningrandarray}}
		
		\pgfmathsetmacro{\outputOffsetX}{11}
		\pgfmathsetmacro{\outputOffsetY}{0.5}
		\pgfmathsetmacro{\outputSize}{0.5}
		\pgfmathsetmacro{\outputBlock}{4}
		
		\coordinate (A) at (0 - 0.5, \outputOffsetY + 0.5);
		\coordinate (B) at (\outputOffsetX -0.5, \outputOffsetY - 1);
		\node [myrect={(A) (B)}, opacity=0.2, fill=trevoryellow, draw=none] (inBlock) { };
		\node [above= -15pt and 0pt of inBlock, scale=0.75] (outputBlockLabel) { Locations $(\latentdata_1, \ldots, \latentdata_{\numTaxa})$ };
		
		\foreach \gmemory in {1,...,\globalsize} {
			
			\pgfmathtruncatemacro{\prev}{\gmemory-1}
			\pgfmathtruncatemacro{\next}{\gmemory+1}
			
			\pgfmathsetmacro{\tmp}{\randarray[\gmemory}
			\xdef\colorvalue{\tmp}
			
			\draw[fill=highcolor!\colorvalue!lowcolor] (\samplewidth * \prev, 0) rectangle (\samplewidth * \gmemory, \sampleheight);
			\draw (\samplewidth * \prev, 0.5 * \sampleheight) -- (\samplewidth * \gmemory, 0.5 * \sampleheight);
		}
		
		\pgfmathsetmacro{\blockSize}{16}
		\pgfmathsetmacro{\I}{4}
		\pgfmathsetmacro{\Iend}{\I + \blockSize}
		\pgfmathsetmacro{\J}{44}
		\pgfmathsetmacro{\Jend}{\J + \blockSize}
		
		\coordinate (A) at (\samplewidth * \I - 0.5 * \samplewidth, -0.5 * \samplewidth);
		\coordinate (B) at (\samplewidth * \Iend + 0.5 * \samplewidth, \sampleheight + 2.5 * \samplewidth);
		\node[myrect={(A) (B)}, opacity=0.3, fill=black] (gmemoryI) {};
		\node[scale=0.5, above = -11pt and 0pt of gmemoryI] (gmemoryILabel) { $i: (I - 1) B + 1, \ldots, I B$ };

		\coordinate (A) at (\samplewidth * \J - 0.5 * \samplewidth, -0.5 * \samplewidth);
		\coordinate (B) at (\samplewidth * \Jend + 0.5 * \samplewidth, \sampleheight + 2.5 * \samplewidth);
		\node[myrect={(A) (B)}, opacity=0.3, fill=black] (gmemoryJ) {};
		\node[scale=0.5, above = -11pt and 0pt of gmemoryJ] (gmemoryJLabel) { $j:  (J - 1) B + 1, \ldots, J B$ };
		
		
		
		\pgfmathsetmacro{\amplitude}{1.5pt}
		
		\draw [thick, decorate, decoration={brace, amplitude=\amplitude, mirror}]
		([yshift=-2pt] gmemoryI.south west) -- ([yshift=-2pt] gmemoryI.south east);
		
		\draw [thick, decorate, decoration={brace, amplitude=\amplitude, mirror}]
		([yshift=-2pt] gmemoryJ.south west) -- ([yshift=-2pt] gmemoryJ.south east);

		\pgfmathsetmacro{\threadOffsetX}{0.5}
		\pgfmathsetmacro{\threadOffsetY}{-5}
		\pgfmathsetmacro{\plate}{0.05}
		\pgfmathsetmacro{\threadWidth}{8}
		
		\coordinate (A) at (\threadOffsetX, \threadOffsetY);
		\coordinate (B) at (\threadOffsetX + \threadWidth, \threadOffsetY + 3.5);
		\node [myrect={(A) (B)}, opacity=0.2, fill=trevorblue] (threadBlock) { };

		\node [left=of threadBlock,rotate=+90, anchor=north] (sharedMemoryLabel) { Shared memory };
		\draw [thick, decorate, decoration={brace, amplitude=10pt, mirror}]
		([xshift=-2pt] threadBlock.north west) -- ([xshift=-2pt] threadBlock.south west);
		
		\pgfmathsetmacro{\localOffsetX}{1}
		\pgfmathsetmacro{\localOffsetY}{-2.25}
		
		\coordinate (A) at (\localOffsetX + \samplewidth * 0 - 0.5 * \samplewidth, \localOffsetY + -0.5 * \samplewidth);
		\coordinate (B) at (\localOffsetX + \samplewidth * \blockSize + 0.5 * \samplewidth, \localOffsetY + \sampleheight + 0.5 * \samplewidth);
		\node[myrect={(A) (B)}, opacity=0.3, fill=black, draw=none] (smemoryI) {};
		
		\foreach \lmemory in {1,...,\blockSize} {
			
			\pgfmathsetmacro{\gmemory}{\I + \lmemory}
			\pgfmathtruncatemacro{\prev}{\lmemory-1}
			\pgfmathtruncatemacro{\next}{\lmemory+1}
			
			\pgfmathsetmacro{\tmp}{\randarray[\gmemory}
			\xdef\colorvalue{\tmp}
			
			\draw[fill=highcolor!\colorvalue!lowcolor]
			(\localOffsetX + \samplewidth * \prev, \localOffsetY + 0) rectangle
			(\localOffsetX + \samplewidth * \lmemory, \localOffsetY + \sampleheight);
			\draw  (\localOffsetX + \samplewidth * \prev, \localOffsetY + 0.5 * \sampleheight)
			-- (\localOffsetX + \samplewidth * \lmemory, \localOffsetY + 0.5 * \sampleheight);
		}
		
		\pgfmathsetmacro{\ia}{\blockSize + 1}
		\pgfmathsetmacro{\ib}{\blockSize + \blockSize}
		\pgfmathsetmacro{\localOffsetX}{\localOffsetX + \samplewidth * 1}
		
		\coordinate (A) at (\localOffsetX + \samplewidth * \blockSize - 0.5 * \samplewidth, \localOffsetY + -0.5 * \samplewidth);
		\coordinate (B) at (\localOffsetX + \samplewidth * 2 * \blockSize + 0.5 * \samplewidth, \localOffsetY + \sampleheight + 0.5 * \samplewidth);
		\node[myrect={(A) (B)}, opacity=0.3, fill=black, draw=none] (smemoryJ) {};
		
		\foreach \lmemory in {\ia,...,\ib} {
			
			\pgfmathsetmacro{\gmemory}{\J + \lmemory - \blockSize}
			\pgfmathtruncatemacro{\prev}{\lmemory-1}
			\pgfmathtruncatemacro{\next}{\lmemory+1}
			
			\pgfmathsetmacro{\tmp}{\randarray[\gmemory}
			\xdef\colorvalue{\tmp}
			
			\draw[fill=highcolor!\colorvalue!lowcolor]
			(\localOffsetX + \samplewidth * \prev, \localOffsetY + 0) rectangle
			(\localOffsetX + \samplewidth * \lmemory, \localOffsetY + \sampleheight);
			\draw  (\localOffsetX + \samplewidth * \prev, \localOffsetY + 0.5 * \sampleheight)
			-- (\localOffsetX + \samplewidth * \lmemory, \localOffsetY + 0.5 * \sampleheight);
		}
		
		\path ([yshift=-6pt] gmemoryI.south) -- node (textI) { $\threadsPerBlock$-thread read } ([yshift=2pt] smemoryI.north);
		\path ([yshift=-6pt] gmemoryJ.south) -- node (textJ) { $\threadsPerBlock$-thread read } ([yshift=2pt] smemoryJ.north);
		\draw [thick, ->] ([yshift=-6pt] gmemoryI.south) -- (textI) -- ([yshift=2pt] smemoryI.north);
		\draw [thick, ->] ([yshift=-6pt] gmemoryJ.south) -- (textJ) -- ([yshift=2pt] smemoryJ.north);
		
		
		\coordinate (A) at (\outputOffsetX - 0.5, \outputOffsetY + 0.5);
		\coordinate (B) at (\outputOffsetX + \outputBlock * \outputSize + 0.5, \outputOffsetY - \outputBlock * \outputSize -0.5);
		\node [myrect={(B) (A)}, opacity=0.2, fill=trevoryellow, draw=none] (outputBlock) { };
		\node [above= -15pt and 0pt of outputBlock, scale=0.75] (outputBlockLabel) { Partial sums $\{ \transformR_{I J} \}$ };
		
		\node [right=of outputBlock,rotate=-90, anchor=north] (globalMemoryLabel) { Global memory };
		\draw [thick, decorate, decoration={brace, amplitude=10pt}]
		([xshift=2pt] outputBlock.north east) -- ([xshift=2pt] outputBlock.south east);
		
		\foreach \o in {1,...,\outputBlock} {
			\pgfmathsetmacro{\op}{\o - 1}
			\foreach \p in {1,...,\outputBlock} {
				\pgfmathsetmacro{\pp}{\p - 1}
				\pgfmathtruncatemacro{\tmp}{random(\maxrand)}
				\xdef\colorvalue{\tmp}
				
				\coordinate (A) at (\outputOffsetX + \op * \outputSize, \outputOffsetY - \pp * \outputSize);
				\coordinate (B) at (\outputOffsetX + \o  * \outputSize, \outputOffsetY - \p  * \outputSize);
				\node[myrect={(A) (B)}, fill=outHighcolor!\colorvalue!outLowcolor] (gout\o gout\p) {};
				
			}
		}
		
		\coordinate (A) at (\outputOffsetX + 2 * \outputSize - 0.5 * \outputwidth, \outputOffsetY - 1 * \outputSize + 0.5 * \outputheight);
		\coordinate (B) at (\outputOffsetX + 3 * \outputSize + 0.5 * \outputwidth, \outputOffsetY - 2 * \outputSize - 0.5 * \outputheight);
		\node[myrect={(A) (B)}, opacity=0.2, fill=black] (goutput) {};

		\pgfmathsetmacro{\outputSize}{\blockSize * \blockSize}
		
		%
		%
		%
		%
		%
		%
		%

		\foreach \i in {1,...,4} {
			
			\coordinate (A) at (\threadOffsetX + \plate * \i, \threadOffsetY - \plate * \i);
			\coordinate (B) at (\threadOffsetX + \plate * \i + \threadWidth, \threadOffsetY - \plate * \i + 2.5);
			\node[myrect={(A) (B)}, opacity=1.0, fill=trevorblue!20] (thread\i) { };
			
		}
		
		\coordinate (A) at (\threadOffsetX + \plate * 3, \threadOffsetY - \plate * 3);
		\coordinate (B) at (\threadOffsetX + \plate * 3 + \threadWidth, \threadOffsetY - \plate * 3 + 2.5);
		\node[myrect={(A) (B)}, draw=none] (thread4) {
			Compute transformation in parallel: \\[1em] 
			%
			%
			%
			%
			%
			$ r_{ij} = \frac{ \left( \distance_{ij} - || \latentdata_i - \latentdata_j || \right)^2 }{ 2 \mdsVariance }
			+ \log  \normalCDF{ \frac{|| \latentdata_i - \latentdata_j ||}{ \mdsSD} } $
			%
			%
		};
		
		
		\node (textThread) at ($([xshift=6pt] thread4.east)!0.45!([xshift=-2pt, yshift=-2pt] gout3gout2.center)$)
		{ \,\,\,\,\, $\threadsPerBlock^2$-thread reduction };
		\draw [thick, ->] ([xshift=6pt] thread4.east) -- (textThread) -- ([xshift=-2pt, yshift=-2pt] gout3gout2.center);
		\node [below right=0cm and -2.8cm of textThread] (reduction) {
			\begin{tikzpicture}[scale=0.5, every node/.style={scale=0.5}]
			\pgfmathsetmacro{\outputheight}{0.17}
			\pgfmathsetmacro{\outputwidth}{0.17}
			
			\foreach \outi in {1,...,\blockSize} {
				\pgfmathsetmacro{\previ}{\outi - 1}
				\pgfmathsetmacro{\nexti}{\outi + 1}
				
				\foreach \outj in {1,...,\blockSize} {
					\pgfmathsetmacro{\prevj}{\outj - 1}
					\pgfmathsetmacro{\nextj}{\outj + 1}
					
					\pgfmathtruncatemacro{\tmp}{random(\maxrand)}
					\xdef\colorvalue{\tmp}
					
					\draw [fill=outHighcolor!\colorvalue!outLowcolor]
					(\outputwidth * \previ, \outputheight * \prevj) rectangle
					(\outputwidth * \outi , \outputheight * \outj );
				}
			}
			
			\coordinate (A) at (0, 0);
			\coordinate (B) at (\outputwidth * \blockSize, \outputheight * \blockSize);
			\node[myrect={(A) (B)}, opacity=0.0] (gsmall) {};
			
			\node[left of=gsmall, node distance=2.3cm, scale=3] { $\sum$ };
			
			\draw [thick, decorate, decoration={brace, amplitude=7.5pt}]
			([xshift=0.2cm] gsmall.north east) -- ([xshift=0.2cm] gsmall.south east);
			
			\pgfmathsetmacro{\offX}{2.9}
			\pgfmathsetmacro{\offY}{3.6}
			\pgfmathsetmacro{\levels}{3}
			\pgfmathsetmacro{\finalCount}{pow(2, \levels - 1)}
			
			\foreach \level in {1,...,\levels} {
				\pgfmathtruncatemacro{\levelCount}{pow(2, \levels - \level)}
				\pgfmathtruncatemacro{\step}{pow(2, \level - 1)}
				\foreach \label in {1,...,\levelCount} {
					\pgfmathtruncatemacro{\niceLabel}{\label - 1}
					\node [] (label\niceLabel level\level)
					at (\offX + \label * \step - 0.5 * \step, \offY + \level * -1) { $+$ };
				}
				
			}
			
			\pgfmathtruncatemacro{\endArrow}{\levels - 1}
			\foreach \level in {1,...,\endArrow} {
				\pgfmathtruncatemacro{\levelCount}{pow(2, \levels - \level)}
				\foreach \label in {1,...,\levelCount} {
					\pgfmathtruncatemacro{\niceLabel}{\label - 1}
					\pgfmathtruncatemacro{\belowLabel}{\niceLabel / 2}
					\pgfmathtruncatemacro{\belowLevel}{\level + 1}
					\draw [->] (label\niceLabel level\level) -- (label\belowLabel level\belowLevel);
				}
				
			}
			
			\end{tikzpicture}
		};
	\end{tikzpicture}
	
	\caption{Massive parallelization strategy for computing the log-likelihood: each working group independently reads two separate batches of latent locations data from global memory, computes location pair specific likelihood contributions in parallel, efficiently adds these contributions in a binary reduction and writes the resulting partial sum to global memory.}
	\label{fig:mp}
\end{figure}
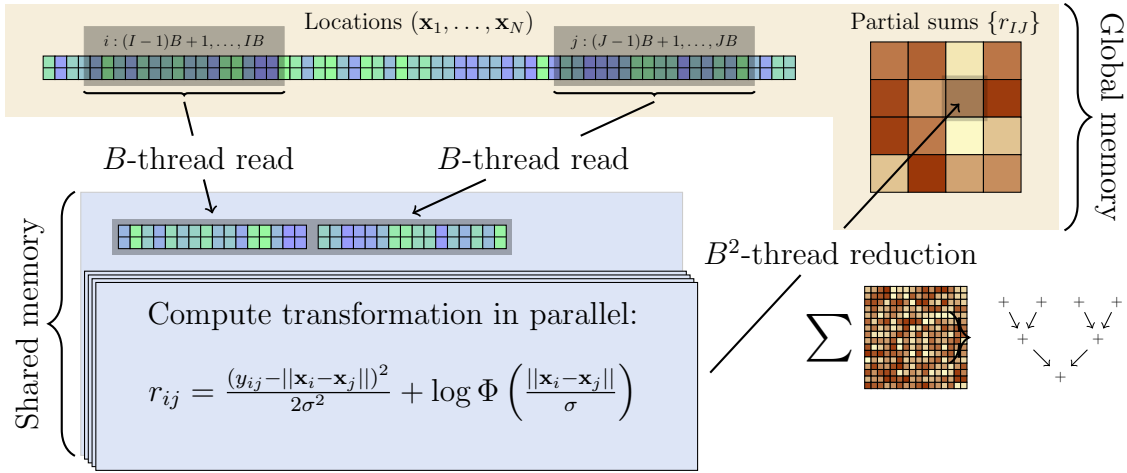

Modern laptop, desktop and server computers hold sockets for 1 to 8 separate CPU chips and each chip consists of 1 to 72 independent processing units called cores that can execute different computing operations simultaneously.
Cores on the same chip share only a small amount of low-level, high-speed cache to facilitate communication, while cores on different chips often share only high-level, slow-speed cache if any.
A drawback of this architecture is that a single memory bus connects the cache to RAM.
	The rate at which data moves across this bus, called the memory bandwidth, is several times less than the total rate of numerical operations across all cores.
	For numerically intensive computation on small amounts of data, this rarely presents a problem.
	As data sizes grow,  memory bandwidth limitations emerge.

Since multi-core hardware is designed to perform independent operations, the operating system provides tools to coordinate their behavior that the computational statistician accesses through software libraries that are built into many programming languages.
\textsc{Threading Building Blocks} (\textsc{TBB}) is one popular, open-source and cross-platform library \citep{Reinders:2007:ITB:1461409} that provides a convenient and expressive application programming interface (API) for multi-core parallelization of transformations and reductions.
For example, the \textsc{R} package \textsc{RcppParallel} \citep{rcppparallel} wraps \textsc{TBB}, making it immediately accessible to \textsc{R} and \textsc{C++} statistical developers.

In a multi-core CPU environment, effective parallelization of the transformation-reductions in Equations (\ref{eq:likelihood}) and (\ref{eq:gradient}) first evenly partitions the task into a modest number of parallel threads $\numThreads$, where $\numThreads \le$ the total number of cores available.  Without SIMD, each thread can only perform the rate-limiting floating point operations one-at-a-time
but benefits from the large size of the CPU caches that automatically hold multiple copies of all of $\latentData$ close to the cores.
Using \textsc{TBB}, we assign elements $\summant_{ij}$ (for the likelihood) or $\gradContribution$ (for the gradient) that hold $j$ constant to the same thread, such that a core loads $\latentdata_{j}$ and $\mathbf{\distance}_j$ (the $j$th column or row of $\distanceMatrix$) into an on-chip register and reuses it many times.
Beyond our specific task partitioning, no specialized programming is necessary for a compiler to generate this code.
In parallel, the threads accumulate $\numThreads$ partial-sums that \textsc{TBB} stores to RAM.
A final, serial reduction of the $\numThreads$ partial-sums takes negligible time.

\paragraph{Many-core GPUs}

GPUs contain 100s to 1000s of cores on a single chip and come, on the smaller side, integrated directly into a CPU or, on the larger side, as add-on cards that interface with laptop, desktop or server computers.
Unlike the independent cores in a CPU, small blocks of GPU-cores must execute the same instructions simultaneously, but on potentially different data.
While this appears to be a strong disadvantage, it greatly simplifies thread management.
Blocks of threads may communicate almost instantly using shared memory in register-space directly in hardware, and efficiently scheduling many more threads to execute than cores hides memory transaction latency with RAM, called global memory on a GPU, because many tasks are in flight simultaneously.
Embarrassingly parallel tasks with no communication and low data reuse, such as independent simulation, run modestly faster on a GPU than CPU because the total rate of numerical operations across 1000s of GPU cores is currently larger than 10s of CPU cores, although the gap is narrowing.
But GPUs have limited memory cache, so exploiting the shared memory with many short-lived cooperative threads leads to the greatest performance boosts.

\newcommand{\blockSize}{B}

To evaluate the log-likelihood transformation-reduction on the GPU, we generate $\numThreads = \numTaxa \times \numTaxa$ threads and task each thread $ij$ with computing only one entry $\summant_{ij}$.
We block threads together in $\blockSize \times \blockSize$ work-groups, such that each work-group $IJ$ contains threads tasked to consecutive
$i$ and consecutive $j$ (Figure \ref{fig:mp}).
By setting $\blockSize = 16$ to a small power-of-two, threads within a work-group can communicate via shared memory.
First, this reduces the number of memory transactions to bring $\latentData$ and $\distanceMatrix$ on-chip by a factor of $\blockSize$.
For example, the first $\blockSize$ threads in each work-group read in corresponding group entries for $\latentdata_{i}$ and the second $\blockSize$ threads read in entries for $\latentdata_{j}$, then all $\blockSize^2$ threads have quick access.
Most importantly, all threads in a work-group independently compute each $\summant_{ij}$ (and concomitant rate-limiting floating point operations) in parallel, for each of the $\blockSize^2$ $ij$ pairs in the work-group. The threads then use shared memory again to perform a binary-tree reduction.
A single thread from each group then writes its partial-sum back to global memory.
A final, serial reduction on the CPU of the $\numTaxa^2 / \blockSize^2$ partial-sums takes negligible time.

Recall that the gradient evaluation consists of $\numTaxa$ independent transformation-reductions, one for the gradient with respect to each $\latentdata_i$.  On the GPU, we generate $S=\numTaxa \times \blockSize$ threads for $\blockSize$ a moderate power of 2 (we choose $\blockSize=128$). In parallel across all $i$, we use $\blockSize$ threads to compute the gradient with respect to each individual $\latentdata_i$.  Each thread uses a for-loop to compute $\lceil \numTaxa / \blockSize \rceil$ gradient contributions $\gradContribution$ across individual $j$s, after which the threads work in concert to perform a comprehensive binary reduction of all the terms contributing to the gradient with respect to $\latentdata_i$. Besides the serial evaluations within each thread, the GPU computes the rate-limiting floating point math operations in a massively parallel manner.
Furthermore, GPUs have high memory bandwidth, so it is not a problem that each thread requires a copy of $\latentdata_i$ and $\mathbf{\distance}_i$.  Finally, every work-group stores its $\latentdata_i$ and $\mathbf{\distance}_i$ in place for efficient reuse.


We write our GPU code in the Open Computing Language (OpenCL), an open-source standard maintained by leading hardware vendors, such as AMD, Apple, IBM, Intel and NVIDIA.
The OpenCL framework allows for a ``program once, execute across many heterogeneous platforms,'' including CPUs, GPUs and other emerging hardware accelerators, portable approach using a familiar C-like syntax.
In OpenCL, we write a single function, called a kernel, for the log-likelihood and for the gradient transformation-reductions, and the library assigns these kernels to each working group independently for parallel evaluation at run-time.

\begin{figure}
	\centering
	\begin{minipage}[t]{0.92\textwidth}
		\begin{minted}[%mathescape,
		linenos,
		numbersep=5pt,
		frame=lines,
		framesep=2mm,autogobble]{nasm}
		movapd (%rax,%rbx,8), %xmm0 ; Load 2 doubles (dp) into register
		subpd  (%rax,%rcx,8), %xmm0 ; Subtract 2 doubles from register
		dppd   $49, %xmm0, %xmm0    ; Take dot-product of register and itself
		sqrtsd %xmm0, %xmm0         ; Square root of 1 double (sd)
		\end{minted}
	\end{minipage}
	\caption{Single instruction, multiple data (SIMD) Intel x86 CPU processor instructions to compute $|| \latentdata_i - \latentdata_j ||$ for $\numTraits = 2$.  These SIMD instructions simultaneously act on 2 double-precision floating-point values.  In 4 lines of code, we approximately halve the total number of instructions executed to compute the distance between two vectors, resulting in almost a 2-fold speed-up.
		\label{fig:simd}}
\end{figure}

\paragraph{Within-core vectorization}

Multi- and many-core processing mainly benefit from the concurrent execution of multiple threads of instructions.
Commonly overlooked in statistical computing stands an alternative form of non-concurrent, data-level parallelism called vector or SIMD processing.
In vector processing, a single instruction directs the core to operate simultaneously on a short vector or packet of data stored consecutively in an extended-length register.
Beginning in the mid 1990s, SIMD processors began arriving in commodity computers.
On Intel x86 hardware, the instruction sets carry the names multiple math extensions (MMX), streaming SIMD extensions (SSE) and, in its most recent form, advanced vector extensions (AVX) that operate on 2 to 8 integer or floating-point values. At the time of writing not widely available, next generation AVX-512 extends AVX from 256 to 512 bit extended registers with availability set to grow over the coming years.

\par

While almost every computer used for statistical computing supports this form of parallelism,
few statistical tools explicitly exploit them, relying on compilers to inject occasional SIMD instructions through their automatic optimization procedures.
Unfortunately, compiler-based automatic loop vectorization remains in its infancy, forcing developers often to hand-code SIMD instructions at bottlenecks.
The learning curve is high and good documentation is scarse, but the performance pay-off makes exploring SIMD worth it.
Expressive libraries wrapped into the \textsc{R} toolchain, like \textsc{RcppXsimd} and \textsc{RcppNT2} \citep{rcppnt2}, are emerging, making SIMD as good as a free-lunch for statistical computing.

One trick to successful SIMD parallelism consists of identifying a rate-limiting transformation in which the input data lie consecutively in memory.   For the log-likelihood and its gradient, the evaluation of $\Phi(\cdot)$ is easily identifiable as the most rate-limiting set of operations using an instruction-level program profiler, such as \textsc{Intel VTune} under Windows and Linux and \textsc{Instruments} on a Mac system.  We attack this bottleneck by calling the required floating-point operations on an entire SIMD extended-length register, as opposed to a solitary floating-value, each time.  Using SSE, we effectively evaluate $\Phi(\cdot)$ over 2 double-precision floating variables at a time.  Using AVX, we effectively evaluate $\Phi(\cdot)$ over 4 doubles at a time and reduce computing time by more than a half.

We illustrate this technique with the calculation of distance between two vectors.
If we physically order in RAM the floating-values of $\latentData$ as $\{x_{11},\ldots,x_{1\numTraits},x_{21},\ldots\}$ -- and pad with $0$ between $\latentdata_{i}$ and $\latentdata_{i+1}$ if $D$ is not even (SSE) or not divisible by 4 (AVX) -- then the transformation $|| \latentdata_i - \latentdata_j || \rightarrow \modelDistance_{ij}$ is ripe for SIMD parallelism.
In the case of SSE, exploiting these in computing the dot-product in $|| \latentdata_i - \latentdata_j ||^2$ approximately halves the number of operations.
Figure \ref{fig:simd} displays the x86 SSE instructions for this transformation when $\numTraits = 2$.
One instruction loads the set of packed doubles (pd) $\{\latentDatum_{i1}, \latentDatum_{i2}\}$ into an extended SIMD register.
The next instruction loads and subtracts $\{\latentDatum_{j1}, \latentDatum_{j2}\}$, leaving
$\{\latentDatum_{i1} - \latentDatum_{j1}, \latentDatum_{i2} - \latentDatum_{j2}\}$ in register.
A third instruction forms the dot-product $(\latentDatum_{i1} - \latentDatum_{j1})^2 + (\latentDatum_{i2} - \latentDatum_{j2})^2$
that is a single double (sd) value and a final non-SIMD instruction returns its square root.
%
%
%
%
%
%
SIMD operations can also lead to super-linear speed-up (${>}X$-fold using $X$-wide SIMD instructions)  because they can be more cache-efficient and better identify data-dependence between instructions.
This latter feature allows modern CPUs to capitalize on instruction-level parallelism through pipelining and out-of-order execution.

We have placed the algorithmic details corresponding to this discussion in Section \ref{sec:algorithms}, where Algorithm \ref{alg:lik} describes our massively parallel implementation of the log-likelihood computations and Algorithm \ref{alg:grad} describes the same for the log-likelihood gradient.

\subsection{Software availability}\label{sec:software}

The Bayesian evolutionary analysis by sampling trees (BEAST) software package \citep{suchard2018bayesian} stands as a popular tool for viral phylogenetic inference.
The package already implements MCMC methods to explore $\density{\sequences, \phylogeneticParameters, \tree}$ under a wide variety of evolution modeling assumptions and \citet{cybis2015assessing} extend BEAST to include $\cdensity{\latentData, \traitVariance}{\tree}$.
Here, we provide an open-source, stand-alone library \textsc{MassiveMDS} \url{http://github.com/suchard-group/MassiveMDS} that efficiently computes $\log \cdensity{\traitData}{\latentData, \mdsVariance}$ and its gradient and currently integrates directly into BEAST via a simple application programming interface (API).
The library contains a combination of C++ code for which standard compilers can generate CPU-specific vectorized instructions at compile-time and OpenCL kernels that the library constructs and compiles at run-time to facilitate GPU-vendor-specific optimization.
Distribution as a stand-alone library source code with a simple API promotes cross-platform compatibility.

To further ease adoption, we have used the \textsc{Rcpp} package \citep{eddelbuettel2011rcpp} to make the library available in the \textsc{R} programming language as a rudimentary package so that \textsc{R} users can  exploit massive parallelization without requiring the tool authors to be experts in parallelization themselves; this design model has served well previously \citep{10.1093/sysbio/syz020}. Finally, we have facilitated
\textsc{R} user access to advanced SIMD tools by making the C++ library \textsc{Xsimd} available with  \textsc{R} wrapper package \textsc{RcppXsimd} \url{http://github.com/OHDSI/RcppXsimd}.

\section{Demonstration}

Each year, seasonal influenza infects at least 10\% of the world population, causing as many as 500,000 deaths.  Prior to 2009, four main influenza subtypes circulated among humans. Of these,  influenza A lineages H3N2 and H1N1 are the most prevalent.  Influenza B subtypes Yamagata and Victoria contribute to decidedly less infections.
\cite{bedford2015global} related this difference in epidemic success to differences in the rate of antigenic evolution.
Indeed, H3N2 and H1N1 have higher rates of `antigenic drift' compared to the less prevalent influenza B counterparts \citep{bedford2014integrating}.
This results in different age-of-infection patterns that coupled with age-dependent air travel intensity explain different migration rates \citep{bedford2015global}.
Antigenic evolutionary rates were estimated using a Bayesian phylogenetic MDS model with antigenic distances arising from costly chemical assays \citep{bedford2014integrating}.  Here, we use a concept of worldwide air traffic space to derive pairwise distances between individual viral samples.
Our goal is to obtain lineage-specific rates of dispersion through this air traffic space using Bayesian phylogenetic MDS.

To this end, massive parallelization facilitates phylogenetic analysis of huge collections of viral data with varied strains. We analyze 1370, 1389, 1393 and 1240 samples of type H1N1, H3N2, VIC and YAM, respectively.   The observed sample originates from 189 different countries, making it ideal for testing the proposed air traffic distance framework.

\subsection{Viral mobility from air traffic}

\begin{figure}[t]
	\centering
	\includegraphics[width=1\textwidth]{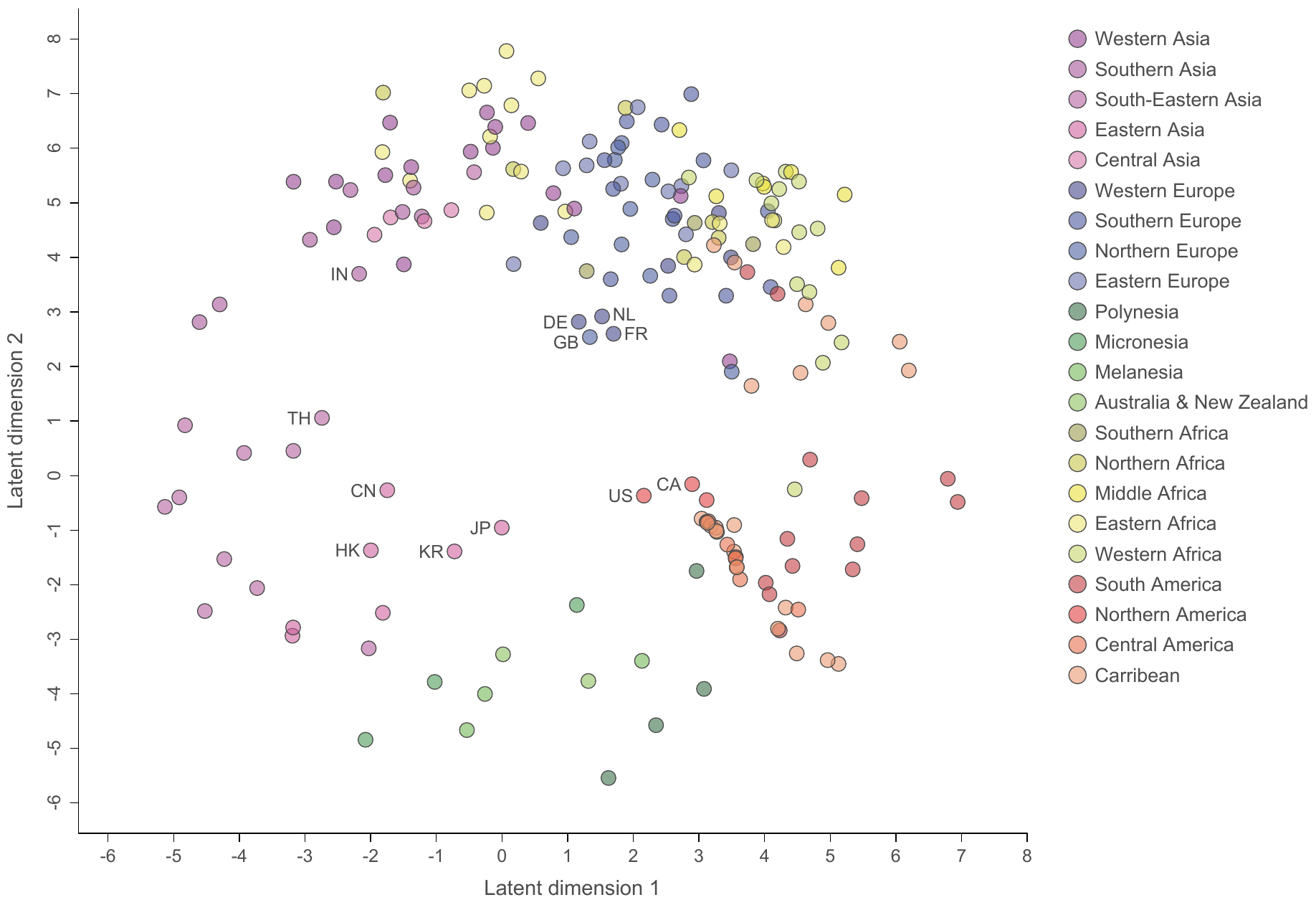}
	\caption{Geography of a worldwide, latent air traffic space.  A 2-dimensional Bayesian multidimensional scaling model with effective worldwide air traffic space distances for data results in 189 country specific posterior medians.}
	\label{fig:geography}
\end{figure}

\begin{figure}[t]
	\centering
	\includegraphics[width=0.7\textwidth]{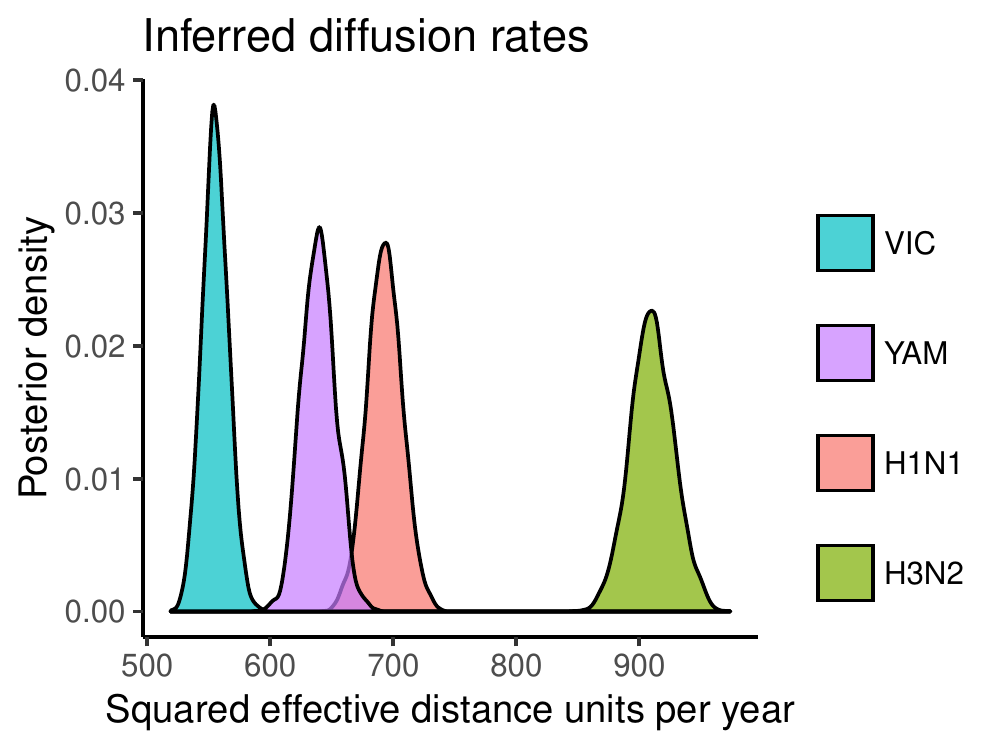}
	\caption{Posterior distributions of strain-specific diffusion rates inferred from 6-dimensional Bayesian phylogenetic multidimensional scaling with effective worldwide air traffic space distances for data.}
	\label{fig:diffusionDensities}
\end{figure}

\newcommand{\assignment}[1]{\phi_{#1}}
\newcommand{\propConstant}{c}
\newcommand{\seats}{s}
\newcommand{\numAirports}{A}
\newcommand{\deff}{d^e}

We use \emph{effective distance} \citep{brockmann2013hidden} between countries to incorporate global transport information into our analysis.
Effective distances summarize global air travel patterns as a network of 4069 nodes (airports) and 25,453 edges (direct connections).  Let $\alpha$ and $\beta$ index two arbitrary nodes on this network.
\citet{brockmann2013hidden} construct $p_{\alpha \beta}$, the probability of traveling from $\alpha$ to $\beta$ based on flight frequency numbers, and use this probability to render
\begin{align*}
\deff_{\alpha \beta} = 1 - \log p_{\alpha \beta}
\end{align*}
the effective distance between the nodes. This measure is inversely proportional to the probability of traveling between nodes, and the log transform guarantees additivity of edge lengths, a direct corollary of the fact that transition probabilities multiply. On the other hand, $\deff_{\alpha \beta}$ does not generally equal $\deff_{\beta \alpha}$ (consider the probability of traveling from New York to the Solomon Islands), so we further symmetrize the measure to make amenable to a continuous latent space representation.  Finally we aggregate the distances by country to form our data.

\begin{figure}
	\centering
	\includegraphics[width=0.7\textwidth]{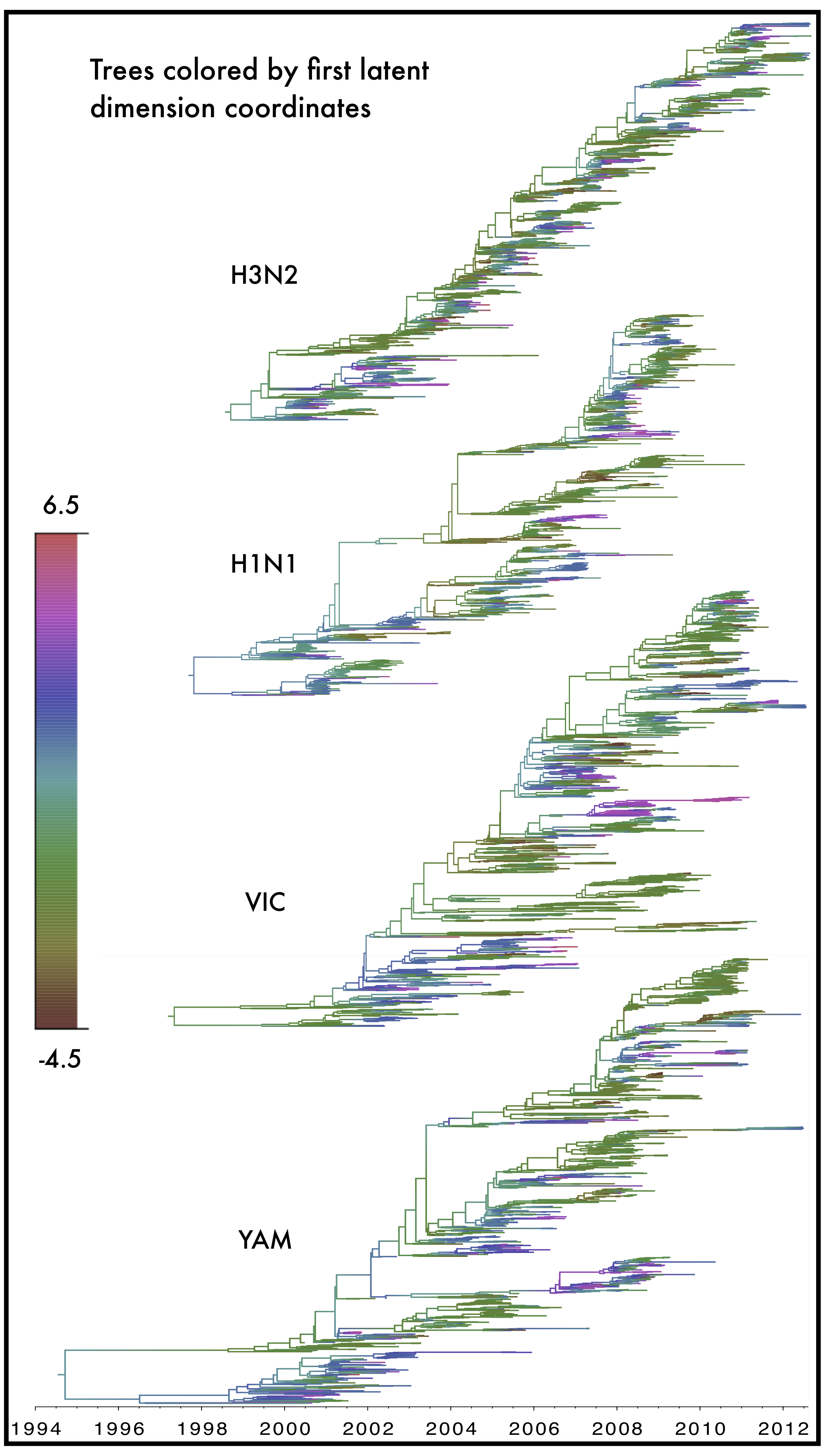}
	\caption{Posterior modes for trees from each strain, colored along first latent dimension of worldwide air traffic space.}\label{fig:treesFigure}
\end{figure}

\subsection{Dispersal inference}\label{sec:DispInf}
\newcommand{\E}{\mbox{E}}

To elicit the prior on the tree-based covariance through time $\treeVariance$ for the latent diffusion process, we incorporate a set of posterior trees from the analysis of \citet{bedford2015global} into the model as a finite mixture.   That analysis assumes the Hasegawa, Kishino and Yano \citep{hasegawa1985dating} process with discretized rate variation \citep{yang1996among} to model sequence substitution as a continuous-time Markov chain along an unknown tree $\tree$.
For a prior distribution over $\tree$, \citet{bedford2015global} elect for a flexible, non-parametric coalescent-based process \citep{gill2013improving} and give each strain its own tree. \citet{bedford2015global} use Haemagglutinin coding sequences, numbering 9139 H3N2, 3789 H1N1, 2577 VIC and 1821 YAM in total and coming from the 9 geographic regions of USA/Canada, South America, Europe, India, North China, South China, Japan/Korea, Southeast Asia and Oceania.


First, we visualize the worldwide air traffic space using a two-dimensional latent space model.
In Figure \ref{fig:geography}, the posterior medians of 189 countries arrange themselves according to our worldwide air traffic distances.
Continental and otherwise geographic blocks of countries (given by similar colors) hew together.
Within blocks, economic powerhouses tend toward the center of the space because they are more closely connected by air travel to other economic giants from other blocks and hence other blocks in general.
For example, the United States (US) `represents' the Americas in red  while Japan (JP) and China (CN) represent East Asia in pink.)

As indicated, the rate of dispersal for each individual viral strain is an important quantity of scientific interest, and a key question is whether one may accurately infer these rates with a phylogenetic MDS model trained on latent airspace data.  For the multivariate Brownian diffusion, tr$(\traitVariance)$ is the univariate measure of instantaneous dispersal satisfying
\begin{align*}
\langle \dx \latentdata,\, \dx \latentdata\rangle = \mbox{tr}\left(\traitVariance\right) \, \dx t \, ,
\end{align*}
where $\traitVariance$ is the same as in Equation \ref{eq:multinormal} and $\dx \latentdata$ is the instantaneous change in $\latentdata$ as described by the stochastic differential equations governing multivariate Brownian motion. To infer these quantities we must choose the latent dimensionality of our MDS model. As described above, we use 5-fold cross-validation, which dictates a 6-dimensional latent space: the average log-likelihoods for latent dimensions 2 through 7 are $-7.1\times 10^{-6}$, $-4.2\times 10^{-6}$, $-3.4\times 10^{-6}$, $-3.5\times 10^{-6}$, $-2.8\times 10^{-6}$, and $-7.0\times 10^{-6}$. 

Conditioning on the finite mixture of trees within a Gibbs sampler, we use a GPU to generate 2 million HMC states in roughly 48 hours. GPU based HMC accelerates sampling over latent locations, but we generate a large number of Markov chain states because changes in tree topology and branch lengths result in an array of posterior geometries, all of which require exploration.  With a fixed tree, one needs to generate an order of magnitude fewer samples for similar problems at similar scale.  Gibbs steps for both the MDS scale parameter $\sigma^2$ and the low-dimensional covariance $\traitVariance$ are straightforward and do not slow sampling of other model parameters.
Using this MCMC sample, we obtain empirical posterior densities for strain-specific dispersal rates and present them in Figure \ref{fig:diffusionDensities}.
In order, the posterior modes of the evolutionary diffusion rates for the four subtypes are 900, 700, 640 and 550 squared effective distance units per year for H3N2, H1N1, VIC, and YAM, respectively.
The relative distributions are in line with discrete migration rate estimates between worldwide regions obtained by \cite{bedford2015global}, and they largely follow differences in posterior means for antigenic drifts for the same lineages (1.01, 0.62, 0.42, and 0.32 for H3N2, H1N1, VIC, and YAM, respectively, \cite{bedford2014integrating}).
That scalings between the studies are different should not be surprising: one set of distances arises from biological measures; another comes from transportation metrics.
Nonetheless, our result corroborates the results of the former study in terms of relative evolutionary rates.

Distinct relative rates of dispersal cohere to qualitatively different phylogenies.
We present posterior modes of inferred strain-specific trees obtained from the same MCMC sample in Figure \ref{fig:treesFigure}.
The trees belonging to the subtype B lineages and H1N1 are much bushier than that of H3N2.
The latter lineage maintains a steady rate of evolution, and the former lineages display a periodic switch between years characterized by long and short branch lengths.
Short branches indicate small effective population sizes -- a result of rapid population turnover -- while long branches indicate large effective population size.
Indeed, \cite{bedford2014integrating} infer similar phylogenies.

Of the four subtypes in our study, YAM has the oldest most recent common ancestor (MRCA), which takes place around 1994.  VIC, the other influenza B subtype, has the next oldest  MRCA (circa 1997), followed by the influenza A subtypes H1N1 (circa 1998) and H3N2 (circa 2000).    Rooted at their MRCAs, the trees extend through worldwide air traffic space: Figure \ref{fig:treesFigure} colors each by position with respect to the first latent dimension.  As a relic of its rapid dispersion, H3N2 has branches that quickly oscillate between brown, green and red as the lineage travels through the latent space.  On the other hand type B viruses (and to a lesser extent H1N1) have entire clusters characterized by a single locality in air traffic space, as indicated by slowly changing hews.



\subsection{Parallelization}\label{sec:ParResults}



\begin{table}
	\centering
	\begin{tabular}{llllllllll}
		\toprule
		Cores &  \multicolumn{2}{c}{1} &  2 &  4 &  6 &  8 & 10 & 12 & \multirow{2}{*}[-4pt]{GPU} \\  \cmidrule{2-3}
		Vectorization & None & SSE & \multicolumn{6}{c}{AVX} &  \\ \cmidrule{4-9}
		Likelihood &  0.41 &  0.71 &  1.98 &  3.80 &  5.57 &  7.29 &  8.55 &  9.31 & 92.25 \\
		Gradient &   0.42 &   0.75 &   1.96 &   3.73 &   5.37 &   7.14 &   8.46 &   9.18& 177.77 \\
		\bottomrule
	\end{tabular}
	\caption{Speedup of graphics processing unit (GPU) and multi-core advanced vector extensions (AVX) computations relative to single core AVX computing.  Single core implementations without single instruction, multiple data (SIMD) and with streaming SIMD extensions (SSE) occupy the bottom left corner.}
	\label{tab:performFig}
\end{table}

To produce the CPU results in this section (as well as Section \ref{sec:scaling}), we use an iMac Pro with a 10-core Intel Xeon processor clocked at 3.0 GHz, 32 GB DDR4 memory (2666 MHz), and 23.75MB cache.  With hyperthreading ($\times2$ instructions per cycle) and AVX ($\times4$), it achieves 240 Gflops peak double-precision floating point performance. For the GPU results, we use an NVIDIA Quadro GP100 with 3485 single-precision floating point CUDA cores and 16 GB HBM2 memory, achieving roughly 5 Tflops double-precision floating point performance.

Table \ref{tab:performFig} compares GPU, multi-core, and single core implementations of log-likelihood and log-likelihood gradient evaluations for 5,338 samples or approximately 14 million pairwise distance data points. We also compare SSE vectorization and no SIMD against AVX vectorization for a single core.  For each processor setting, we perform 100 evaluations and report the average speedup. Reported speedups are relative to AVX-based, single-core processing for each evaluation type (likelihood or gradient).


Single core, SSE computations are slightly slower (likelihood: 590 ms; gradient: 950 ms), and non-vectorized computations are significantly slower (likelihood: 1,016 ms; gradient: 1,715 ms), than single core, AVX CPU computations.  For multi-core CPU processing, we find relative speedups that scale roughly linearly with the number of cores.  With 12 cores, AVX averages 44 ms per likelihood evaluation and 77 ms per gradient evaluation, roughly 10 times faster than the respective 420 and 716 ms per evaluation for single core AVX.  Again, these results arise from an application with 5338 locations, so they are particularly encouraging given that we use a CPU with roughly 24MB cache, maxing out at $\numTaxa\approx1730$ 6-dimensional locations ($\latentData_{\numTaxa\times6}$) and their pairwise observations $(\distanceMatrix_{\numTaxa^2}$) stored in double-precision.  Nonetheless, we posit that a top-of-the-line, modern CPU with 70MB cache capable of holding roughly 2955 locations and concomitant observation matrix could deliver even greater speed.

Averaging 4.5 ms for the likelihood and 4 ms for the gradient, GPU implementations are reliably around 100 times faster than single core, AVX implementations.  For inference for the illustration with H1N1, H3N2, VIC, and YAM, the GPU requires 48 hours to generate 2 million HMC states.  Back of the envelope calculation shows the same posterior inference requiring almost a full solar revolution for the single core AVX implementation.
We place additional scaling studies in Section \ref{sec:scaling} of the Appendix.

Finally, we allow that there are many criteria by which to judge software and respective hardware implementations.  The NVIDIA Quadro GP100 we use is top-of-the-line and typically represents a  purchase additional to whichever computer one might be working with, whereas the majority of CPUs do not. We also recognize that such technology advances at great speeds, gradually becoming less expensive and proliferating in use. For these competing reasons we have developed software to exploit the strengths of both CPUs and GPUs, whether through vectorized, multi-core or many-core processing.

\section{Discussion}

We developed Bayesian phylogenetic MDS to visualize pathogen diffusions and learn related scientific quantities.  We used `airspace distance' between viral samples to model the dispersion
of four different strains of flu: H1N1, H3N2, Victoria and Yamagata. Doing so, we obtained established strain-specific diffusion rates.

But inference for large collections of viral samples is not easy. We showed that Bayesian MDS is ripe for parallel computation, and that massive parallelization provides massive speedups for likelihood evaluations, likelihood-gradient evaluations and, hence, HMC iterations.  In particular, GPU-based calculations were over 100 times faster than respective single-core based calculations and over 20 times faster than respective multi-core calculations.  In practical terms, massive parallelization can finish in a day what a single core can do in a year!  Moreover, these massive accelerations are available to Bayesian MDS in general and not limited to phylogenetic MDS.

We note that there are other models that are worth exploring:  \citet{hoff2002latent} outlines latent space approaches that are alternatives to MDS; \citet{ramsay1982some} provides alternatives to the truncated normal such as, e.g. the inverse-Gaussian; \citet{oh2007model} employ a mixture of Gaussians as prior over latent positions.  All three of these directions would be amenable to phylogenetic extensions similar to that of MDS developed here.  Indeed, a phylogenetic extension of \citet{oh2007model} would be useful for clustering pathogens, and, hence, predicting evolutionary dynamics.   For viral samples accompanied by metadata labels, one might use latent locations as predictors of, e.g., patient outcomes.  In this case, \citet{holbrook2017bayesian} provides a road map for joint inference over the hierarchical model's MDS and predictive components.

A different kind of question is whether one might make GPU and multi-core SIMD speedups available for a broader class of Bayesian models.  \citet{li2019neural} use neural networks to approximate an arbitrary model's log-posterior gradient and thus avoid expensive HMC gradient computations in a Big Data setting.  On the other hand, GPUs greatly accelerate fitting and evaluation of deep neural networks \citep{bergstra2011theano}.  It seems natural to combine these insights to power HMC based Bayesian inference on a massive scale.

Less straightforward are geometric extensions to phylogenetic MDS.  For example, \citet{zhou2018hyperbolic} rely on the similarities between hyperbolic space and tree space as defined in \citet{billera01} (i.e. negative curvature) to visualize tree structure using the Poincar\'e ball.  Inference for a respective Bayesian model could be done using an intrinsic version of geodesic Monte Carlo \citep{holbrook2018geodesic}.  Another interesting, geometrically inspired model is Lorentzian MDS \citep{clough2017embedding}.  Here, time between samples would contribute negative distance while space between sequences would contribute positive distance, leading to visualization with non-symmetric axes.
Geometric and otherwise, all the above directions are potentially fruitful for Bayesian phylogenetic inference.


\section*{Acknowledgments}

The research leading to these results has received funding from the European Research Council under the European Union's Horizon 2020 research and innovation programme (grant agreement no.~725422-ReservoirDOCS) and from  the National Institutes of Health (R01 AI107034, R01 HG006139 and LM011827) and the National Science Foundation (IIS 1251151 and DMS 1264153).
We gratefully acknowledge support from NVIDIA Corporation with the donation of parallel computing resources used for this research.
The Artic Network receives funding from the Wellcome Trust through project 206298/Z/17/Z.
PL acknowledges support by the Research Foundation -- Flanders (`Fonds voor Wetenschappelijk Onderzoek -- Vlaanderen', G066215N, G0D5117N and G0B9317N).
GB acknowledges support from the Interne Fondsen KU Leuven / Internal Funds KU Leuven under grant agreement C14/18/094.

\newpage
\appendix

\section{Algorithms}\label{sec:algorithms}

We present Algorithms \ref{alg:lik} and \ref{alg:grad} for parallel computing of the likelihood and log-likelihood gradient, respectively.   Algorithmic details remain the same for multi-core CPU and GPU approaches, but implementations do not.  Also, for the CPU implementation, $B$ is the size of the SIMD extended register, but it is the size of the work group for the GPU implementation.

\begin{algorithm}[H]
	\caption{Parallel computation of likelihood} \label{alg:lik}
	\begin{algorithmic}[1]
		\ParFor{$IJ\in \{1,\dots,\lfloor N/B\rfloor\}\times  \{1,\dots,\lfloor N/B\rfloor\}$}
		\ParFor{$ij \in \{1,\dots,B\} \times  \{1,\dots,B\}$}
		\If{$I\times B +i <N$ \textbf{and} $J\times B + j<N$}
		\State copy $\latentdata_i$, $\latentdata_j$ to local \Comment{first $2B$ threads}
		\State calculate $\delta_{ij}$  \Comment{all $B^2$ threads, using SIMD Figure \ref{fig:simd}}
		\State copy $\distance_{ij} $ to local
		\State locally compute $r_{ij}$
		\EndIf
		\EndParFor
		\State compute partial sum $r_{IJ}$  \Comment{binary tree reduction on chip}
		\State write $r_{IJ}$ to global memory \Comment{using single thread}
		\EndParFor
		\State $\cdensity{\traitData}{\latentData, \mdsVariance} \gets \sigma^{N(1-N)/2} \exp \left(-\sum_{IJ}r_{IJ} \right)$ \Comment{on CPU}
	\end{algorithmic}
\end{algorithm}

\begin{algorithm}[H]
	\caption{Parallel computation of gradient} \label{alg:grad}
	\begin{algorithmic}[1]
		\ParFor{$i \in \{1,\dots,N\}$}
		\State copy $\latentdata_i$ to local \Comment{$B$ threads}
		\ParFor{$J\in \{1,\dots,\lfloor N/B\rfloor\}$}
		\State $j \gets J$
		\While{$j < N$}
		\State copy $\latentdata_j$ to local \Comment{$B$ threads}
		\State $\Delta_{ij} \gets \latentdata_i - \latentdata_j$ \Comment{first two steps of SIMD Figure \ref{fig:simd}}
		\State calculate $\delta_{ij}$  \Comment{final two steps of SIMD Figure \ref{fig:simd}}
		\State copy $\distance_{ij} $ to local
		\State $\nabla_{iJ} \gets \nabla_{iJ}  -\left( \frac{(\delta_{ij}-\distance_{ij})}{\mdsVariance} +\frac{\phi(\delta_{ij}/\sigma)}{\sigma \Phi(\delta_{ij}/\sigma)} \right) \frac{\Delta_{ij}}{ \delta_{ij}} $
		\State $j \gets j + B$
		\EndWhile
		\EndParFor
		\State $\frac{\partial}{\partial \latentdata_i} \log \cdensity{\traitData}{\latentData, \mdsVariance} \gets \sum_J \nabla_{iJ} $   \Comment{binary tree reduction on chip}
		\EndParFor
	\end{algorithmic}
\end{algorithm}

\section{Additional scaling studies}\label{sec:scaling}
\begin{figure}[t]
	\centering
	\includegraphics[width=0.8\textwidth]{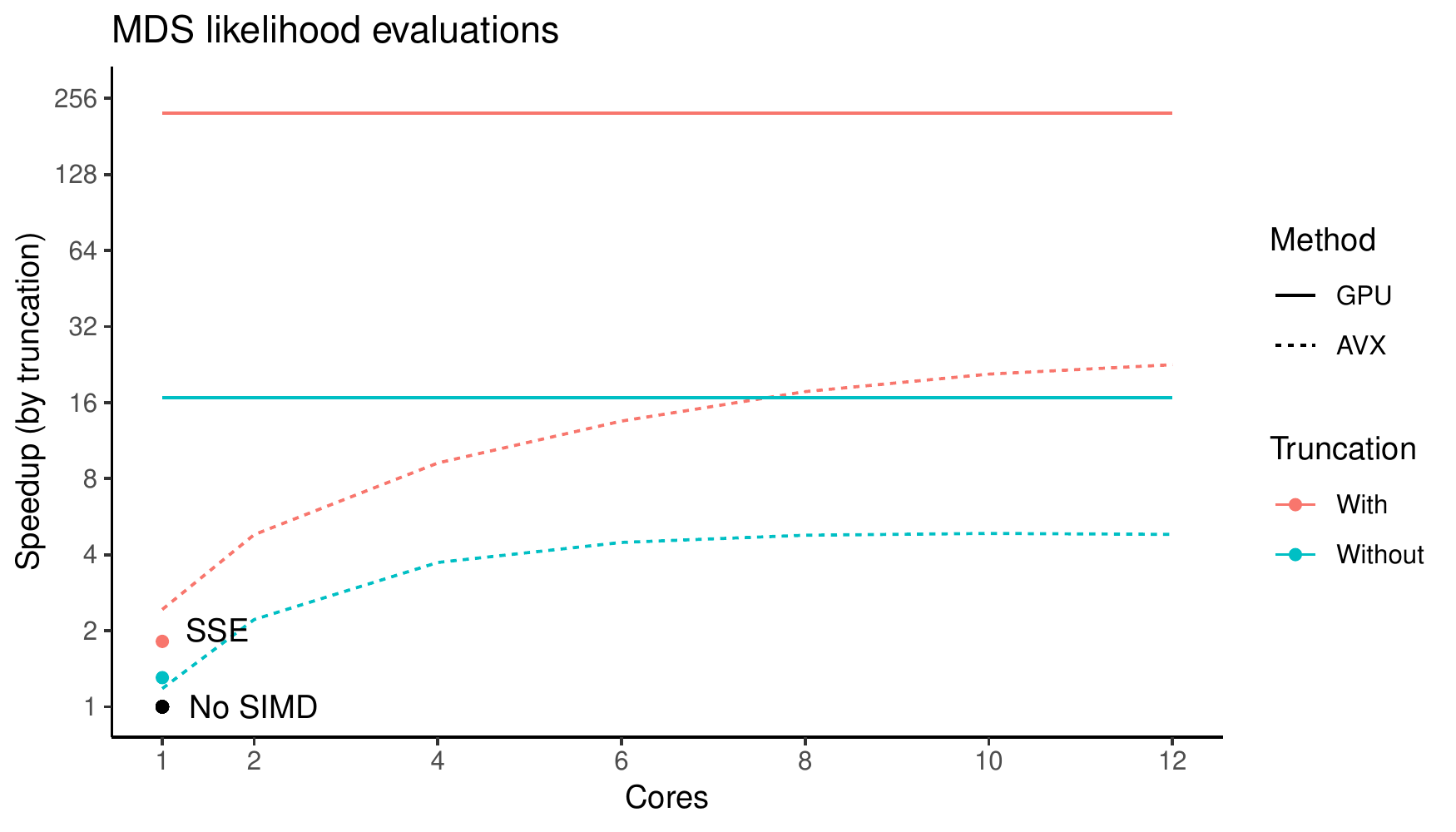}
	\caption{Speedup of graphics processing unit (GPU) and multi-core advanced vector extensions (AVX) computations over single core implementations of multi dimensional scaling (MDS) likelihood with and without truncation. No single instruction, multiple data (No SIMD; baseline, black) implementation and streaming SIMD extensions (SSE; colored) occupy the bottom left corner.}
	\label{fig:truncFig}
\end{figure}

\begin{figure}[t]
	\includegraphics[width=\textwidth]{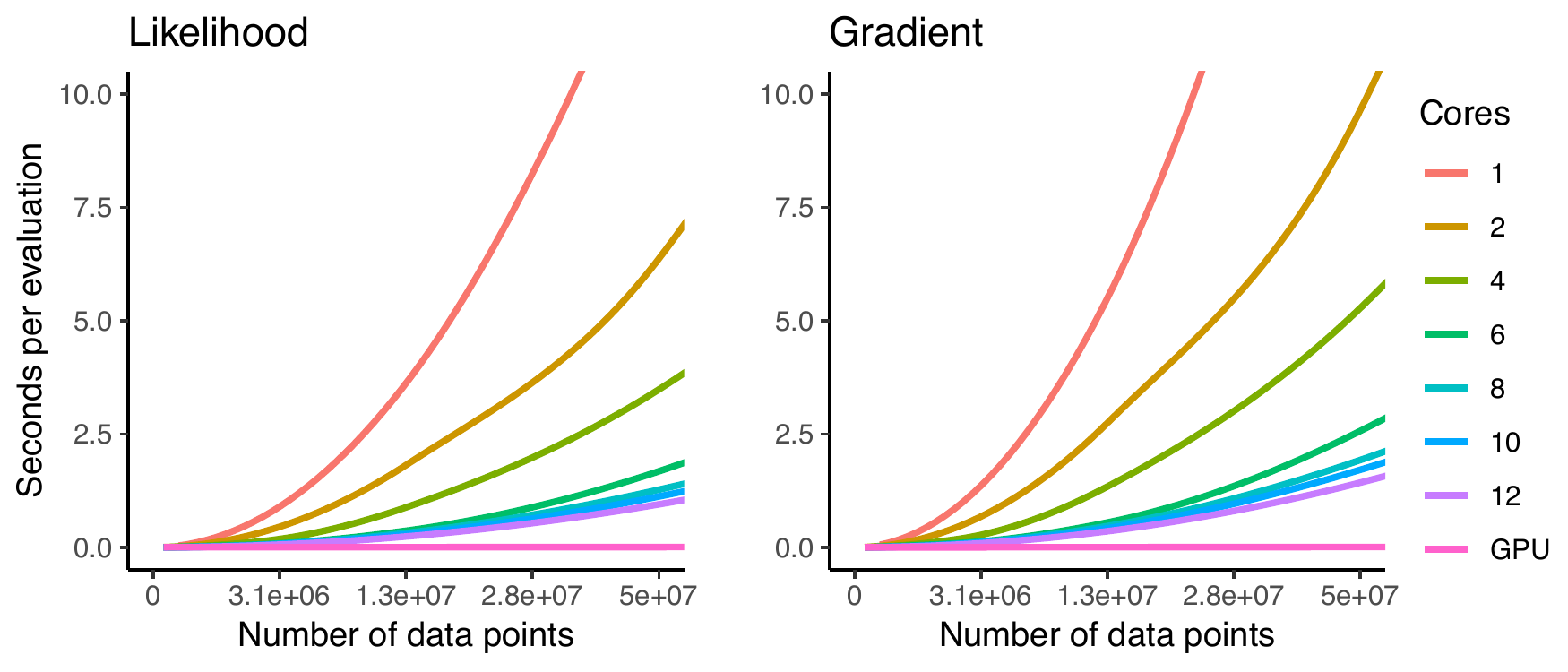}
	\caption{Seconds to evaluate likelihood and gradient using central processing unit (CPU) and graphics processing unit (GPU) as a function of data size. For both likelihood and gradient, computation time grows exponentially for CPU and logarithmically for GPU implementation. }
	\label{fig:time_by_N}
\end{figure}

First, we provide insight into one of the main computational challenges of the MDS likelihood: the truncation term
\begin{align*}
\sum_{i<j}\log  \Phi \left(\delta_{ij} / \sigma \right) \, .
\end{align*}
This term is computationally intensive because of its $\order{N^2}$ floating point operations. Since the term is the sum of a single simple function applied independently to all $\delta_{ij}$, parallelization should deliver significant speedups.
Figure \ref{fig:truncFig} shows relative speedups (over no SIMD, single core) of likelihood computation using SIMD vector processing, multi-core CPU and GPU processing. In 100 independent iterations, we generate 5,338 samples (approximately 14 million data points) and time the likelihood and gradient evaluations.    When the truncation term is not calculated, the 12-core implementation is only 4 times faster than the single core without SIMD, and GPU calculations are only 16 times faster.  But when truncation is included (i.e., the correct model), 12-core implementation is more than 16 times faster, and GPU 200 times faster, than the single core implementation without SIMD.

Figure \ref{fig:time_by_N} shows seconds per likelihood and log-likelihood gradient evaluations for GPU and multi-core implementations.  Results are based on the two-dimensional latent space model and distances arising from randomly sampled Gaussian points. Speeds are averaged over 100 independent tests.  Lower values correspond to less computing time.  Both for the likelihood and the gradient, GPU evaluation speed (bottom) stays orders of magnitude faster than multi-core and single core evaluation speeds.

\end{document}